\documentclass[aps,prd,twocolumn,groupedaddress]{revtex4-1}
\usepackage{graphicx}
\include{amssym}

\begin{document}
\title{$1/N_c$ Nambu -- Jona-Lasinio model: $\pi^0$, $\eta$ and $\eta'$ mesons}
\author{A.\,A.\,Osipov\footnote{Email address: osipov@nu.jinr.ru}}
\affiliation{Joint Institute for Nuclear Research, Bogoliubov Laboratory of Theoretical Physics, 141980 Dubna, Russia}

\begin{abstract}
We continue to study the properties of the light pseudoscalar nonet within the combined framework of Nambu -- Jona-Lasinio model and $1/N_c$ expansion, assuming that current quark masses count of order $\mathcal O(1/N_c)$. The masses, mixing angles and decay constants of the $\pi^0$, $\eta$ and $\eta'$ are calculated. The role of the $U(1)_A$ anomaly is emphasized. It is shown that the gluon anomaly suppresses the leading order effects that might otherwise be substantial for the $\eta\to 3\pi$ amplitude. A detailed comparison with the known results of $1/N_c$ chiral perturbation theory is made. 
\end{abstract}

\maketitle

\section{Introduction}
In the world of massless up, down and strange quarks, the Lagrangian of quantum chromodynamics (QCD) is symmetric under $U(3)_L\times U(3)_R$ chiral transformations. This symmetry, however, is violated spontaneously by the non-zero quark condensate, and by the axial $U(1)_A$ anomaly. The response of the quark-gluon vacuum to the spontaneous symmetry breaking is the excitation of eight Goldstone modes $\pi$, $K$ and $\eta$, while the ninth Goldstone mode $\eta'$ receives a large mass due to the $U(1)_A$ anomaly. In the real world, the $\pi$, $K$ and $\eta$ mesons acquire their masses because chiral $SU(3)_L\times SU(3)_R$ symmetry is broken explicitly by the non-zero quark masses $m_u\neq m_d\neq m_s$. As a consequence, the physics of the pseudo Goldstone bosons is based on three pillars: the value of the quark condensate, the strong $U(1)_A$ anomaly, and the pattern of the light quark masses. 

These three essential elements of pseudoscalars dynamics are deeply correlated. Indeed, if $U(1)_A$ were a good symmetry in nature, one would have a light isoscalar particle $L$ with the mass $m^2_L\leq 3m_\pi^2$ \cite{Weinberg:75}. Moreover, if the ratio $(m_d -m_u)/(m_d+m_u)\simeq 0.3$ were appreciable, i.e., if it were not hidden by the peculiar features of chiral dynamics indicated above, the isotopic spin symmetry would be substantially violated so that the mass eigenstates of neutral pseudoscalar mesons would be pure, each containing only one quark flavor pair: $\bar uu$, $\bar dd$, and $\bar ss$ \cite{Gross:79}. Another manifestation of the correlation is the surprisingly large mass of pseudoscalars compared with the light quark masses. As we learned from current algebra, the masses of pseudoscalars are proportional to the current quark masses. In the case of the pion, the formula reads $m_\pi^2=B(m_u+m_d)$, where the constant $B$ is non-zero in the chiral limit $B_0=-\langle \bar qq\rangle_0/F^2$. The quark condensate and the pion decay constant imply a very large factor $B_0\simeq 2.5\,\mbox{GeV}$ (we give here the estimate obtained in the framework of the Nambu and Jona-Lasinio (NJL) model, chiral perturbation theory gives a relatively smaller, but still pretty large value $B_0\simeq 1.4\,\mbox{GeV}$) which 
significantly enhances the effect of light quark masses.

The consequences of explicit and spontaneous violation of chiral symmetry are most interestingly reflected in the physical properties of neutral pseudoscalars $\pi^0$, $\eta$ and $\eta'$. 
It is this aspect of chiral dynamics that this article is devoted to. The $\pi^\pm$ and $K$ mesons were considered in our previous work \cite{Osipov:23}. The study is based on the effective meson Lagrangian originated from the effective $U(3)_L\times U(3)_R$ symmetric four-quark interactions of the NJL type \cite{Nambu:61a,Nambu:61b}, where we, following Leutwyler's idea \cite{Leutwyler:96a,Leutwyler:96b}, count the light quark masses to be of order $\mathcal{O}(1/N_c)$. 

To succeed in the quantitative description of effects related to the explicit chiral symmetry breaking, we use the asymptotic expansion of the quark determinant \cite{Osipov:21a,Osipov:21b,Osipov:21c}, which is based on the Fock-Schwinger proper time method and the Volterra series. This powerfull tool allows not only to isolate divergent parts of quark loop diagrams, but also accurately reproduce their flavor structure. The latter circumstance is fundamental in studying the explicit violation of chiral symmetry in the NJL model.   

A huge number of papers have been devoted to the study of the properties of the neutral pseudo Goldstone particles. Therefore, we consider it necessary at the beginning of our presentation to answer the question of what is the novelty of the results presented here in comparison with the already well-known achievements in this actively developed area \cite{Gan:22}.

In answering this question one should stress that the NJL model has not previously been used for the theoretical study of neutral pseudoscalar states under the assumption that $m_i=\mathcal{O}(1/N_c)$ (except for a short letter \cite{Osipov:22b}). We think that the implementation of this idea may allow us to look at the results of $1/N_c$ chiral perturbation theory ($1/N_c\chi$PT) \cite{Leutwyler:96a,Leutwyler:96b,Taron:97,Kaiser:00,Goity:02,Bickert:20} from a new angle, since the prospect opens up to directly relate the low-energy constants of $1/N_c\chi$PT with the parameters of the NJL model, i.e., with the main characteristics of the hadronic vacuum.

The hypothesis $m_i=\mathcal{O}(1/N_c)$ should not be taken literally, i.e., as a direct use of the Taylor expansion in powers of quark masses. It is well known that the chiral series is not of the Taylor type. It contains non-analytic terms, so-called chiral logarithms. The assumption to count the light quark masses of order $1/N_c$ shifts the contributions of chiral logarithms to the next-to-next-to leading order (NNLO). Correspondingly, at the next, NNNLO, step, it is necessary to take into account the contribution of two-loop meson diagrams, and so on. Thus, a full account of current quark masses by the naive summation of the Taylor series is a misleading procedure, because it does not account for essential contributions of chiral logarithms arising at higher powers of light quark masses. Hence only the leading order (LO) result and the first $1/N_c$ correction (NLO) to it has a polynomial form in the current quark masses. It is this approximation that is used here to study the $\pi^0$-$\eta$-$\eta'$ system.

Our paper also reports on some progress in describing explicit chiral symmetry breaking in comparison with previous schemes developed on the basis of the NJL model \cite{Volkov:84,Wadia:85,Volkov:86,Ebert:86,Osipov:92, Bijnens:93,Osipov:13}. In particular, we show a deep connection between the results obtained here and similar results known from the $1/N_c\chi$PT (previous NJL approaches have been less successful in this).

But there are also differences. In the NJL model, the kinetic term of the free meson Lagrangian is the result of calculating the self-energy meson diagram with a quark loop. We show that this leads to a redefinition of the original meson fields collected in the matrix $U=e^{i\phi} \in U(3)$. As a result, the neutral states in the octet-singlet basis $\phi_a$ $(a=0,3,8)$ are not pure, namely $\phi_a=\sum_b F^{-1}_{ab}\phi^R_b$ is a superposition of rescaled eigenfunctions $\phi_a^R$ for which the kinetic part of the Lagrangian is diagonal. The appearance of such impurities in $\phi_a$ is the result of the explicit violation of chiral symmetry. This has physical consequences: the $U(1)_A$ anomaly contributes at the next-to-leading order (NLO) to the masses of $\pi^0$, $\eta$, and $\eta'$ mesons suppressing effects of flavor and isospin symmetry breaking \cite{Osipov:23b}.

One example of such suppression is found in the calculation of the $\eta$-$\pi^0$ mixing angle $\epsilon$. It is known that the interference with $\eta'$, in the LO of the $1/N_c$ expansion, strongly affects the amplitude of the $\eta\to 3\pi$ decay which is proportional to $\epsilon$. This effect was discussed by Leutwyler \cite{Leutwyler:96b}, who pointed out on its similarity with the other effect occurring in the mass spectrum of $\eta$-$\eta'$. He has found that chiral symmetry implies that the same combination of effective coupling constants which determines the small deviation from the Gell-Mann-Okubo formula also specifies the symmetry breaking effects in the decay amplitude and thus ensures that these are small. Indeed, below we show that the NLO correction significantly suppresses the isospin symmetry breaking effect observed at the LO. As a result, one can not only obtain the phenomenological values of the $\eta$ and $\eta'$ masses, but also reduce the isospin breaking angle $\epsilon$ to the value established early in \cite{Gross:79}.   

We find a second example of the suppression effect of the gluon anomaly calculating the $\eta$-$\eta'$ mixing angle. It is known that in $1/N_c\chi$PT this angle is dramatically reduced to about $-10^\circ$ from its LO value of $-18.6^\circ$ \cite{Goity:02}. We show that in the $1/N_c$ NJL model the magnitude of the NLO corrections is rather small: its LO value $-15^\circ$ is corrected to $-15.8^\circ$ after the NLO contributions are taken into account. 

We also consider the scheme with two mixing angles, which is widely discussed in the literature \cite{Schechter:93,Moussallam:95,Feldmann:98,Kroll:05,Escribano:05}, and demonstrate that in the NJL model it arises, as an approximation, after the NLO corrections are included. Unfortunately, within the framework of $1/N_c$ NJL model, we fail to find a rigorous theoretical justification for this mixing scheme. The latter is possible only in the presence of off-diagonal terms in the kinetic part of the free meson Lagrangian, which explicitly violates Zweig's rule \cite{Schechter:93}. Since quark loop diagrams contributing to the self-energy meson graph satisfy this rule, off-diagonal vertices do not arise in the NJL model. Nonetheless, we show that the scheme with one mixing angle guarantees the fulfillment of the well-known relations between weak decay constants \cite{Leutwyler:98}.

The article is organized as follows. In Sec.\,\ref{s2}, we present the form of the free Lagrangian for the $\eta$-$\eta'$-$\pi^0$ fields, which arises as a result of the asymptotic expansion of the quark determinant. Additionally, the contributions of the gluon anomaly and the interaction that violates the Okubo-Zweig-Iizuka (OZI) rule are considered. In Sec.\,\ref{s3}, we calculate the coupling constants $f_0$, $f_3$ and $f_8$ of neutral pseudoscalars and discuss their connection with already known results. In Sec.\,\ref{s4},  the masses and mixing angles are calculated. In Sec.\,\ref{s5},  we consider the octet-singlet basis and calculate the weak decay constants. In particular, it is detailed here how the NLO corrections effectively lead to the well-known scheme with two mixing angles. The physical content of the initial fields $\phi_a$ is discussed in Sec.\,\ref{s6}. In Sec.\,\ref{s7}, we shortly discuss the strange-nonstrange mixing scheme. The comments about regularization dependence of our results are given in Sec.\,\ref{s8}. The latter may be useful in order to have some idea of the theoretical uncertainties behind the results presented in the paper. Our conclusions are collected in Sec.\,\ref{s9}. In order not to clutter up the text with technical details, we put them in three Appendixes.

\section{Basic elements}
\label{s2}
Let us first make a remark regarding the original form of four-quark interactions considered in this and our previous paper \cite{Osipov:23}. They include the $U(3)_L\times U(3)_R$ chiral invariant terms describing the scalar, pseudoscalar, vector and axial-vector nonets. This reflects the the symmetry of QCD at $N_c\to\infty$. Actually this set of four-quark interaction channels is not complete in the spirit of Fierz transformations. The symmetry of massless QCD allows the Fierz-invariant terms, describing the singlet vector and axial-vector four-quark couplings \cite{Klimt:90}. We neglect them here as it has also been done in \cite{Klimt:90} when describing the meson spectrum. The reason is that singlet-octet degeneracy is quite well realized in the empirical mass data on spin-1 mesons. Theoretically, it can be understood as an indication that corresponding couplings are $1/N_c$ suppressed in comparison with the channels considered in our paper.    

To study the main characteristics of the neutral pseudoscalars (masses, mixing angles and decay constants) we need only a part of the effective Lagrangian describing non-interacting $\pi^0$, $\eta$ and $\eta'$ fields. Recall that in the NJL model the effective meson Lagrangian results from the evaluation of the one-loop quark diagrams. On the one hand, this requires a redefinition of the initial field functions, and on the other hand, it allows one to calculate the meson coupling constants appearing as a result of such a redefinition. The details of the one-loop calculations have been presented in our previous work \cite{Osipov:23}, so let us write out only the final result arising for the diagonal pure flavor states of the pseudoscalar nonet, $\phi_i$  $(i=u,d,s)$,
\begin{equation}
\label{lagrneutr}
{\mathcal L}_{\phi^2}=\!\!\! \sum_{i=u,d,s}\!\left[\frac{\kappa_{Aii}}{16G_V}(\partial_\mu\phi_i)^2-\frac{M_i m_i}{4G_S}  \phi_i^2\right]. 
\end{equation} 
Here the coupling constants $G_S$ and $G_V$ characterize the strength of the $U (3)_L \times U (3)_R$ chiral symmetric four-quark interactions with spin zero and one correspondingly. Their dimension is (mass)$^{-2}$, and at large $N_c$ they are of order $\mathcal O(1/N_c)$. $M_i$ is the mass of the constituent i-quark. The heavy constituent masses arise through the dynamic breaking of chiral symmetry and are related to the masses of light quarks $m_i$ by the gap equation. The diagonal elements of the matrix $\kappa_{A}$ is obtained in the result of eliminating the mixing between pseudoscalar and axial-vector states. They can be expressed through the main parameters of the NJL model 
\begin{equation}
\label{kappa}
(\kappa_{A})_{ii}^{-1}=1+\frac{\pi^2}{N_c G_V M_i^2 J_1(M_i)},
\end{equation}   
where
\begin{equation}
J_1(M)=\ln\left(1+\frac{\Lambda^2}{M^2}\right)-\frac{\Lambda^2}{\Lambda^2+M^2}.
\end{equation}
Here, $\Lambda$ is the cutoff characterizing the scale of spontaneous symmetry breaking. The values of the parameters were fixed in \cite{Osipov:23}. We collect them in Table \ref{ParameterSets}. 

As one can see from (\ref{lagrneutr}), the quark one-loop diagrams generating the kinetic part of the free Lagrangian lead to a diagonal quadratic form in the flavor basis, which, after redefining the fields 
\begin{equation}
\phi_i = \sqrt{\frac{4G_V}{\kappa_{Aii}}} \phi^R_i \equiv \frac{\phi_i^R}{f_i},
\end{equation}
takes the conventional form
\begin{equation}
\label{kinpart}
\frac{1}{4}\sum_{i=u,d,s}(\partial_\mu\phi_i^R)^2=\frac{1}{2}\sum_{a=0,3,8}(\partial_\mu\phi_a^R)^2.
\end{equation}
The new field $\phi_i^R$ has the dimension of mass because the coupling  constant $f_i$ has this dimension and the initial field $\phi_i$ is a dimensionless quantity. The transition from the flavor components $\phi_i^R$ to the octet-singlet ones $\phi_a^R$ is described by the matrix $O$ given by Eq.\,(\ref{O}). Due to this transformation, the transition to the octet-singlet basis $\phi_a^R$ does not destroy the diagonal form of Eq.\,(\ref{kinpart}). As a consequence, the unscaled field $\phi_a$ has admixture of scaled components  $\phi_b^R$, with $b\neq a$ and vice versa $\phi_a^R=\sum_b F_{ab}\phi_b$ (see Appendix \ref{app1} for details). The non-diagonal elements of the symmetric matrix $F_{ab}$ given by Eq.\,(\ref{elementsF}) violate flavor symmetry. Both the $SU(3)$ breaking term $F_{08}$ and isospin breaking terms $F_{03}$, $F_{38}$ are $1/N_c$-suppressed. A bit later we will dwell on the connection of the elements $F_{ab}$ with the decay constants of pseudoscalars.
 
\begin{table*}
\caption{The six parameters of the model $\Lambda$, $G_{S}$, $G_{V}$, $m_{u}$, $m_{d}$, and $m_{s}$ are fixed by using the meson masses $m_{\pi^0}$, $m_{\pi^+}$, $m_{K^0}$, $m_{K^+}$, the pion decay constant $f_\pi$ and the cutoff $\Lambda$ as an input. The electromagnetic corrections to the masses of charged mesons are estimated taking into account the violation of Dashen's theorem at next to leading order in $1/N_c$. To do this, we additionally used the value of $f_K$ and the phenomenological data on the $\eta\to 3\pi$ decay rate. All units, except dimensionless quantities $\delta_M$, $a$, $[\Lambda ]=\text{GeV}$, and $\left[G_{S,V}\right]=\text{GeV}^{-2}$, are given in MeV.}
\label{ParameterSets}
\begin{footnotesize}
\begin{tabular*}{\textwidth}{@{\extracolsep{\fill}}cccccccccccccccc@{}}
\hline
\hline 
\multicolumn{1}{c}{$\Lambda$} 
& \multicolumn{1}{c}{$G_S$}
& \multicolumn{1}{c}{$G_V$}
& \multicolumn{1}{c}{$m_u$}
& \multicolumn{1}{c}{$m_d$}
& \multicolumn{1}{c}{$m_s$}
& \multicolumn{1}{c}{$M_0$}
& \multicolumn{1}{c}{$-\langle\bar qq\rangle^{1/3}_0$}
& \multicolumn{1}{c}{$M_u$}
& \multicolumn{1}{c}{$M_d$}
& \multicolumn{1}{c}{$M_s$}
& \multicolumn{1}{c}{$F$}
& \multicolumn{1}{c}{$f_\pi$}
& \multicolumn{1}{c}{$f_K$}  
& \multicolumn{1}{c}{$\delta_M$}
& \multicolumn{1}{c}{$a$}\\
\hline
 $1.1$ 
& $6.6$  
&  $7.4$ 
& $2.6$ 
& $4.6$ 
& $84$ 
& $274$ 
& $275$ 
& $283$   
& $290$  
& $567$    
& $90.5$  
& $92.2$  
& $111$ 
& $ 0.67$  
& $3.50$ \\
\hline
\hline
\end{tabular*}
\end{footnotesize} 
\end{table*}

For the mass term in Eq.\,(\ref{lagrneutr}) we find   
\begin{equation}
-\mathcal L^{\mbox{\footnotesize m}}_{\phi^2} =\!\!\!\!\sum_{i=u,d,s}\!\!\! \frac{M_im_i(\phi_i^R)^2}{4G_Sf_i^2} \! = \frac{1}{2}\!\! \sum_{a=0,3,8}\!\!\! \phi_a^R \mathcal M_{1ab}^2\phi_b^R,
\end{equation}
where $\mathcal M_{1}^2$ is a symmetric matrix with the elements
\begin{eqnarray}
\label{mass1}  
  &&(\mathcal M_1^2)_{00}=\frac{1}{3G_S}\left(u_u^2+u_d^2+u_s^2\right), \nonumber \\
  &&(\mathcal M_1^2)_{88}=\frac{1}{6G_S} \left(u_u^2+u_d^2+4u_s^2\right), \nonumber \\
  &&(\mathcal M_1^2)_{33}=\frac{1}{2G_S} \left(u_u^2+u_d^2\right), \nonumber \\
  &&(\mathcal M_1^2)_{08}=\frac{1}{3\sqrt 2 G_S} \left(u_u^2+u_d^2-2u_s^2\right), \nonumber \\
  &&(\mathcal M_1^2)_{03}=\frac{1}{\sqrt 6 G_S} \left(u_u^2-u_d^2\right), \nonumber \\
  &&(\mathcal M_1^2)_{38}=\frac{1}{2 \sqrt 3 G_S} \left(u_u^2-u_d^2\right).
\end{eqnarray}  
Here and below, for the convenience of writing formulas, we use the notation 
\begin{equation}
\label{uin}
\frac{M_im_i}{(f_i)^n}\equiv u^n_i.
\end{equation}

Now it is necessary to take into account two important points -- the $U(1)_A$ anomaly and the violation of the OZI rule -- both explained within the $1/N_c$ expansion \cite{Hooft:74,Veneziano:79,Witten:79,Witten:79b, Witten:80,Veneziano:80}. The $U(1)_A$ anomaly contributes to pseudoscalar masses given by Eqs.\,(\ref{mass1}) already at leading order (notice that we count $m_i$ to be of order $\mathcal O(1/N_c)$). The OZI-violating interactions are responsible for the $1/N_c$ correction to the leading order result. The Lagrangians corresponding to these processes have the form of the product of two traces. At the quark-gluon level, such a contribution comes from diagrams with quark loops coupled through the pure gluon exchange. 

The Lagrangian breaking the $U(1)_A$ symmetry was obtained in \cite{Veneziano:80,Trahern:80,Ohta:80,Ohta:81}. Using this result, we set
\begin{equation}
\label{U1}
\mathcal L_{U(1)}=\frac{\lambda_U}{48}\left[\mbox{tr}\left(\ln U-\ln U^\dagger \right) \right]^2 
=-\frac{\lambda_U}{2}\phi_0^2,
\end{equation}
where $U= e^{i\phi }$, and $\phi =\sum_r\phi_r \lambda_r$, $r = 0,1,\ldots ,8$, the matrix $\lambda_0 = \sqrt{2/3}$ and $\lambda_1,\ldots ,\lambda_8$ are the eight Gell-Mann matrices of $SU(3)$.
The dimensional constant $\lambda_U=\mathcal O (N_c^0)$ is the topological susceptibility of the purely gluonic theory, $[\lambda_U ]=M^4$. 

This Lagrangian implies the following contributions to the matrix elements of the $\eta'$-$\eta$-$\pi^0$ mass matrix 
\begin{eqnarray}
\label{mass2}
(\mathcal M^2_2)_{00}&=&\frac{\lambda_U}{f_0^2}, \nonumber\\
(\mathcal M^2_2)_{08}&=&\frac{\sqrt{2}\lambda_U}{3f_0}\left(  \frac{1}{f_3}- \frac{1}{f_s} \right), \nonumber \\
(\mathcal M^2_2)_{03}&=&\frac{\lambda_U}{\sqrt{6}f_0} \left(  \frac{1}{f_u}-\frac{1}{f_d} \right),
\end{eqnarray}
where the couplings $f_0$ and $f_3$ are given in Eqs.\,(\ref{f083}).  

Notice that, because of the Eq.\,(\ref{mixphi}), an additional mixing is induced between the rescaled neutral components, which is associated with violations of the isospin and $SU(3)_f$ symmetries beyond the leading order. Here only the terms which are responsible for leading and next-to-leading order contributions (in $1/N_c$ counting) are retained. 

The Lagrangian violating the OZI rule has the form \cite{Leutwyler:96a}
\begin{equation}
\label{OZI}
\mathcal L_{OZI}=\frac{i\lambda_{Z}}{\sqrt 6}\mbox{tr}(\phi )\mbox{tr}\left[\chi\left(U^\dagger-U \right)\right],
\end{equation}
where $\lambda_{Z}=\mathcal O(N_c^0)$ is a dimensional constant $[\lambda_{Z}]=M^2$ and $\chi$ is given by the diagonal matrix 
\begin{equation}
\chi =\frac{1}{G_S}\mbox{diag} \left(u_u^2, u_d^2, u_s^2 \right).
\end{equation}
As we will see, the counting rule $\lambda_{Z}\sim N_c^0$ leads to a coherent picture for the masses and decay constants of the pseudoscalar nonet. 

The quadratic part of the Lagrangian (\ref{OZI}) is 
\begin{equation}
\label{OZI-M}
\mathcal L_{OZI}\to  \frac{2\lambda_{Z}}{G_S}\phi_0 \!\!\sum_{i=u,d,s} \!\! u_i^2 \phi_i
\end{equation} 
and contributes only at next to leading order $1/N_c^2$ in the matrix elements
\begin{eqnarray}
\label{mass3}
(\mathcal M^2_3)_{00}&=&-\frac{4\sqrt 2\lambda_Z}{\sqrt 3 G_S f_0} \sum_{i=u,d,s} \!\! u_i^3, \nonumber \\
(\mathcal M^2_3)_{08}&=&-\frac{2\lambda_Z}{\sqrt 3 G_Sf_0} \left(u_u^3+u_d^3-2u_s^3\right), \nonumber \\
(\mathcal M^2_3)_{03}&=&-\frac{2\lambda_Z}{G_Sf_0}\left(u_u^3-u_d^3\right), 
\end{eqnarray}
which interfere with the next-to-leading order contribution of the gluon anomaly.
 
\section{Decay constants $f_i$}
\label{s3}
Our next task is to isolate the leading order contribution together with the first $1/N_c$ correction to it in the formulas above. Here we realize this plan for the decay couplings of neutral states
\begin{equation}
\label{fi}
f_i=\sqrt{\frac{\kappa_{Aii}}{4G_V}}.
\end{equation}
For that we need the nontrivial solution of the gap equation $M_i(m_i)$. The latter, in the considered approximation, can be written as a sum \cite{Osipov:23}  
\begin{equation}
\label{expm}  
M_i(m_i)=M_0+M'(0) \, m_i +\mathcal O (m_i^2),
\end{equation}  
where $M_0$ is the mass of the constituent quark in the chiral limit $m_i\to 0$, and  
\begin{equation}
\label{Mprime}
M'(0)=\frac{\pi^2}{N_cG_S M_0^2 J_1(M_0)}\equiv a.
\end{equation}

Then, from Eq.\,(\ref{fi}), Eqs.\,(\ref{f083}) and (\ref{elementsF}) we find
\begin{eqnarray} 
\label{f038ex}
f_0&=&F\left( 1+(2\hat m+m_s)\frac{a-\delta_M}{6M_0}\right)=F_{00}, \nonumber \\
f_8&=&F\left( 1+(\hat m+2m_s)\frac{a-\delta_M}{6M_0}\right)=F_{88}, \nonumber \\
f_3&=&F\left( 1+\hat m \frac{a-\delta_M}{2M_0}\right)=f_\pi =F_{33}, 
\end{eqnarray}
where $\hat m=(m_u+m_d)/2$ and 
\begin{equation}
a-\delta_M=2a(1-\kappa_{A0})\left[1-\frac{\Lambda^4 J_1(M_0)^{-1}}{(\Lambda^2+M_0^2)^2}\right].
\end{equation}
Here $F$ and $\kappa_{A0}$ are the values of the pion decay constant $f_\pi$ and $\kappa_{Aii}$ at $m_i=0$. 

Further, according to Leutwyler \cite{Leutwyler:98}, $1/N_c\chi $PT provides the relation among the decay constants:
\begin{equation}
\label{LR}
f_8^2=\frac{4}{3} f_K^2-\frac{1}{3} f_\pi^2,
\end{equation}
which is valid to first nonleading order. It can easily be verified that, on using Eq.\,(\ref{f038ex}) and expressions for the kaons decay couplings obtained in \cite{Osipov:23}
\begin{equation}
f_K\equiv \frac{f_{K^\pm}+f_{K^0}}{2}=F\left( 1+(\hat m+m_s)\frac{a-\delta_M}{4M_0}\right),
\end{equation} 
relation (\ref{LR}) is also satisfied in our approach. 

The other relation which is a direct consequence of the approach developed here is
\begin{equation}
\label{FK}
f_0^2=\frac{2}{3} f_K^2+\frac{1}{3} f_\pi^2.
\end{equation} 
This result is known from \cite{Feldmann:98}, where the authors used a different method to obtain it.   

For the constants $f_0$ and $f_8$, in addition to the above quadratic relations, one can establish the following linear relations with the constants $f_{\pi}$ and $f_{K}$ 
\begin{equation}
f_0=\frac{1}{3} \left(2f_K+f_\pi\right), \quad f_8=\frac{1}{3} \left(4f_K-f_\pi\right).  
\end{equation}
In contrast to the formulas (\ref{LR}) and (\ref{FK}), there are no higher-order terms in these relations which must be systematically discarded.   

The non-diagonal elements of the matrix $F_{ab}$ in the considered approximation are 
\begin{eqnarray}
F_{38}&\!=\!&\frac{F(m_u\!-\! m_d)}{4\sqrt{3}M_0}(a\!-\!\delta_M)\!=\!-\frac{1}{\sqrt{3}}(f_{K^0}\!-\!f_{K^\pm} ),\nonumber \\
F_{03}&\!=\!&\frac{F(m_u\!-\! m_d)}{2\sqrt{6}M_0}(a\!-\!\delta_M)\!=\!-\sqrt{\frac{2}{3}}(f_{K^0}\!-\!f_{K^\pm} ), \nonumber \\
F_{08}&\!=\!&\frac{F(\hat m\!-\!m_s)}{3\sqrt{2}M_0}(a\!-\!\delta_M)\!=\!-\frac{2\sqrt{2}}{3}(f_K\!-\!f_\pi ).
\end{eqnarray}
They are negative. The first two are associated with the isospin symmetry breaking, and the last one with the violation of $SU(3)$ symmetry: $F_{08}/F_{03}=2R/\sqrt{3}$, where $R=(m_s-\hat m)/(m_d-m_u)$.

\section{Masses and mixing angles}
\label{s4}
Let us now  consider meson mass relations. For that we expand the elements of the resulting mass matrix $\mathcal M^2=\sum_{i=1,2,3} \mathcal M^2_i$ (see Eqs.\,(\ref{mass1}), (\ref{mass2}) and (\ref{mass3})) in powers of $1/N_c$ retaining only the first two terms.
\begin{equation}
\label{massmatrix}
\mathcal M^2_{ab}= \mu^2_{ab}+\Delta \mu^2_{ab}+\mathcal O(1/N_c^3).
\end{equation} 

The leading $1/N_c$-order result is
\begin{eqnarray}
\label{loeta}
\mu^2_{00}&=&\frac{2}{3}B_0\left(2\hat m+m_s\right)+\lambda_\eta^2,  \nonumber \\
\mu^2_{88}&=&\frac{2}{3}B_0\left(\hat m+2m_s\right), \nonumber \\  
\mu^2_{33}&=&2B_0 \hat m =\bar \mu_{\pi^{\pm}}^2, \nonumber \\
\mu^2_{08}&=&-2\frac{\sqrt 2}{3}B_0\left(m_s-\hat m\right), \nonumber \\
\mu^2_{03}&=&-\sqrt{\frac{2}{3}}B_0\left(m_d-m_u\right), \nonumber \\
\mu^2_{38}&=&-\frac{1}{\sqrt 3}B_0\left(m_d-m_u\right),
\end{eqnarray}
where $\lambda^2_\eta \equiv \lambda_U/F^2$, and 
\begin{equation}
B_0=\frac{2G_VM_0}{G_S\kappa_{A0}}=\frac{M_0}{2G_SF^2}=-\frac{\langle\bar qq\rangle_0}{F^2}.
\end{equation}
It coincides with the formulas obtained by Leutwyler \cite{Leutwyler:96a}, but in the case under consideration, all parameters (except for $\lambda_U$) are related to the main parameters of the four-quark dynamics.

It is clear from (\ref{loeta}), that mixing of $\phi_3^R$ with $\phi_0^R$ and $\phi_8^R$ is due to the breaking of isospin symmetry. In the first order in the mass difference $m_d -m_u$, this mixing is removed by rotating to small angles $\epsilon'$ and $\epsilon$, respectively. The $\phi_0^R $-$ \phi_8^R$ mixing is due to the breaking of $SU(3)$ symmetry and can be removed by rotating to the angle $\theta$. With an accuracy to the first order in the breaking of isospin symmetry, the transformation of the neutral components to the physical $\pi^0$, $\eta$, and $\eta'$ states has the form
\begin{eqnarray}
\label{ortho}
\phi_0^R&=& \pi^0 \left(\epsilon' \cos\theta -\epsilon\sin\theta\right)-\eta \sin\theta +\eta'\cos\theta,  \nonumber\\
\phi_8^R&=&\pi^0 \left(\epsilon\cos\theta +\epsilon'\sin\theta\right)+\eta \cos\theta +\eta'\sin\theta, \nonumber \\
\phi_3^R&=&\pi^0-\epsilon\eta-\epsilon'\eta'.
\end{eqnarray}
This orthogonal transformation diagonalizes the mass matrix $\mathcal M^2$, giving the eigenvalues (the mass squares) and eigenvectors of physical states (for more details see Appendix \ref{app2}).

The result of the diagonalization of the mass-matrix (\ref{loeta}) is well known: The predicted mass of the $\eta$ meson $m_\eta = 494\,\mbox{MeV}$ is much smaller than its phenomenological value $m_\eta = 548\,\mbox{MeV}$ and the angle $\theta$ is $\theta \simeq -20^\circ$. The numerical values of the parameters used are given in Table \ref{table2} (see set (a)). 

Recall that the difference between the masses of the charged and neutral pions is due primarily to the electromagnetic interaction. The contribution of the strong interaction is proportional to $(m_d-m_u )^2$ and is thereby negligibly small. The model estimate is
\begin{equation}
m_{\pi^0}^2=\bar m_{\pi^\pm}^2 +\frac{B_0}{2M_0}(m_d-m_u)^2\delta_M\simeq \bar m_{\pi^\pm}^2.
\end{equation}
Here, the overline indicates that the masses were obtained without taking into account electromagnetic corrections.

\begin{table*}
\caption{In the first row, set $(a)$, we show the leading order result for mixing angles $\theta$, $\epsilon$ and $\epsilon'$. Quark masses are given in MeV, $\theta$ in degrees, small angles $\epsilon$ and $\epsilon'$ in radians. The numerical value of $\lambda_\eta^2$ (in $\mbox{GeV}^2$) is extracted from the experimental value of $m_{\eta'}$. Set $(b)$ describes a fit which takes into account the first $1/N_c$-correction. In this case the light quark masses are given up to NLO corrections included. The two input parameters $\lambda_\eta^2$ and $\Delta_N$, are fixed by the phenomenological masses of $m_{\eta}$, $m_{\eta'}$.}
\label{table2}
\begin{footnotesize}
\begin{tabular*}{\textwidth}{@{\extracolsep{\fill}}lcccccccccccccc@{}}
\hline
\hline 
\multicolumn{1}{c}{Set}
& \multicolumn{1}{c}{$m_u$}
& \multicolumn{1}{c}{$m_d$}
& \multicolumn{1}{c}{$m_s$}
& \multicolumn{1}{c}{$\lambda_\eta^2$}
& \multicolumn{1}{c}{$\Delta_N$}
& \multicolumn{1}{c}{$\theta$}
& \multicolumn{1}{c}{$\theta_0$}
& \multicolumn{1}{c}{$\Delta\theta$}
& \multicolumn{1}{c}{$\epsilon$}
& \multicolumn{1}{c}{$\epsilon_0$}
& \multicolumn{1}{c}{$\Delta\epsilon$}  
& \multicolumn{1}{c}{$\epsilon'$} 
& \multicolumn{1}{c}{$\epsilon_0'$} 
& \multicolumn{1}{c}{$\Delta\epsilon'$}\\
\hline
$(a)$ 
& $2.6$ 
& $4.6$ 
& $93$ 
& $0.671$ 
& $-$ 
& $-$   
& $-19.7^{\circ}$  
& $-$    
& $-$  
& $0.0187$  
& $-$ 
& $-$
& $0.0033$ 
& $-$  \\
$(b)$  
& $2.6$ 
& $4.6$ 
& $84$ 
& $0.805$ 
& $0.36$ 
& $-15.76^\circ$   
& $-14.97^\circ$  
& $-0.79^\circ$    
& $0.0114$  
& $0.0177$  
& $-0.0063$ 
& $0.0021$
& $0.0033$ 
& $-0.0012$ \\
\hline
\hline
\end{tabular*}
\end{footnotesize} 
\end{table*}

Now, we do the next step and calculate the first correction $\Delta\mu_{ab}^2$ to the leading term. This correction includes the contributions from Eqs.\,(\ref{mass1}), (\ref{mass2}) and (\ref{mass3}). Note that when calculating these corrections, we systematically neglect the terms $(m_d-m_u)^2$, replacing, for example, the sum $m_u^2+m_d^2 =2\hat m^2 +(m_d-m_u)^2/2$ only by its first term. This is also in agreement with the accuracy with which the rotation matrix (\ref{ortho}) is defined.
As a result, we find
\begin{eqnarray}
\label{DM}  
\Delta \mu_{00}^2&=&\frac{2B_0}{3}\left[(2\hat m^2+m_s^2)\frac{\delta_M}{M_0}-2\Delta_N(2\hat m+m_s)\right], \nonumber\\
\Delta \mu^2_{08}&=& \frac{2\sqrt 2}{3}B_0 (m_s-\hat m) \left[ \Delta_N-(m_s+\hat m)\frac{\delta_M}{M_0}\right], \nonumber \\
\Delta \mu^2_{88}&=& \frac{2B_0}{3M_0}(\hat m^2+2m_s^2) \delta_M, \nonumber \\ 
\Delta \mu^2_{33}&=&\frac{2B_0}{M_0} \hat m^2\delta_M, \nonumber \\
\Delta\mu^2_{03}&=&\sqrt{\frac{2}{3}}B_0 (m_d\!-\!m_u)\!\left(\Delta_N-2\hat m\frac{\delta_M}{M_0}\right)
, \nonumber \\
\Delta \mu^2_{38}&=& -\frac{1}{\sqrt 3}\frac{B_0}{M_0}(m_d^2\!-\!m_u^2) \delta_M,
\end{eqnarray}  
where
\begin{equation}
\label{dN}
\Delta_N\equiv 2\sqrt{6}\frac{\lambda_Z}{F^2}+\lambda_UG_S\frac{a-\delta_M}{2M_0^2}.
\end{equation}

Here it is appropriate to make a few remarks about formula (\ref{DM}). Let us first comment on the origin of different contributions. Corrections caused by Eq.\,(\ref{mass1}) are the terms that remain after we put $\Delta_N=0$. In fact, they coincide up to a common factor with the known result of $1/N_c\chi$PT  \cite{Goity:02}. The correspondence between factors is 
\begin{equation}
\frac{\delta_M}{M_0}\leftrightarrow 16\frac{B_0}{F^2} \left(2L_8^r-L_5^r\right).
\end{equation}  

Next, the corrections proportional to $\lambda_U$ in (\ref{DM}) are related with the $U(1)_A$ anomaly: The corresponding contribution to $\Delta\mu_{00}^2$ arises due to the NLO correction to the coupling $f_0$ in (\ref{mass2}). The other two contributions to $\Delta\mu_{08}^2$ and $\Delta\mu_{03}^2$ are the result of an admixture of rescaled neutral octet components in the singlet field $\phi_0$ described by the Eq.\,(\ref{mixphi}). Both account for the symmetry breaking corrections due to the $U(1)_A$ anomaly. Such corrections interfere with the OZI violating contributions of Lagrangian (\ref{OZI}) and, as a result, the effective coupling constant $\Delta_N$ arises. If we compare the formulas for $\Delta\mu_{08}^2$, $\Delta\mu_{03}^2$ with the analogous expressions obtained in \cite{Goity:02}, one can establish a correspondence
\begin{equation}
\label{dNcomp}
\Delta_N\leftrightarrow - \rho/2 =\Lambda_1/2 -\Lambda_2+4L_5M_0^2/F_0^2,
\end{equation} 
where on the right-hand side we have retained the notation of work \cite{Goity:02}, so one should not confuse the notation $M_0$ (singlet mass) adopted there with the constituent quark mass $M_0$ used here. The only difference between the NJL approach considered here and the $1/N_c\chi$PT is the absence of the NLO term $-M_0^2\Lambda_1$ in our expression for $\Delta\mu_{00}^2$. Probably it is this circumstance that leads to different estimates for the mixing angle $\theta$ in the compared approaches.

The representation of the mass matrix $\mathcal M^2$ as the sum of the leading contribution and the $1/N_c$ correction to it implies a similar representation for all parameters of the transformation that is used to diagonalize the mass matrix: $\theta=\theta_0+\Delta\theta$, $\epsilon =\epsilon_0+\Delta\epsilon$, $\epsilon' =\epsilon_0'+\Delta\epsilon'$. Accordingly, the eigenvalues obtained should have a similar form (see Appendix \ref{app2} for details).  

To obtain numerical values, we fix the main parameters of the model as it was in the case of the charged particles (see the Table \ref{ParameterSets}). Additionally, the phenomenological values of the masses of $\eta$ and $\eta'$ mesons are used to fix the topological susceptibility $\lambda_U$ and the OZI-violating coupling constant $\lambda_Z$. As a result, we obtain the values of the mixing angles $\theta$, $\epsilon$ and $\epsilon'$ (see set (b) in Table \ref{table2}). 

Numerically, the $\eta-\eta'$ mixing angle $\theta =-15.8^\circ$ predicted by the model is consistent with a recent result from lattice QCD $\theta =(-15.1^{+5.9}_{-6})^\circ$ \cite{Kordov:21}, and phenomenology: (a) $\theta =(-15.4 \pm 1.0)^\circ $ \cite{Feldmann:98}; (b) $\theta =(-16.9 \pm 1.7)^\circ $ (this value was deduced from the rich set of $J/\psi$ decays into a vector and a pseudoscalar meson) \cite{Bramon:97}; (c) $\theta =(-15.5 \pm 1.3)^\circ $ (this is a result of thorough analysis of many different decay channels in which the authors took into account the flavor $SU(3)$-breaking corrections due to constituent quark mass differences) \cite{Bramon:99}. 

It should be also noted that the angle $\theta$ obtained here differs noticeably from the estimate $\theta\simeq -10^\circ$ worked out in the framework of $1/N_c\chi$PT \cite{Goity:02}. We have already pointed out the reason for this discrepancy above. Here we note that the $1/N_c$ NJL model does not lead to a huge effect from taking into account NLO contributions observed in \cite{Goity:02}. As one can see from the Table\,\ref{table2}, the LO result $\theta_0$ receives only a 5\% NLO correction. 

Numerical estimates show that the mixing angles $\epsilon$ and $\epsilon'$ are substantially modified at NLO. The corrections account for around $35\%$ of the LO result. In particular, the mixing angle $\epsilon$ is found to be $\epsilon =0.65^\circ$ while the LO result is $\epsilon_0=1.0^\circ$. This result can be compared with the estimate $\epsilon =0.56^\circ$ that arises in $\chi$PT when only octet degrees of freedom are included \cite{Gasser:85}. The similar behaviour is found for the angle $\epsilon'=0.12^\circ$ which is equal $\epsilon_0'=0.19^\circ$ at LO. 

Since the NJL model in the LO reproduces analytically the mixing angles $\epsilon$ and $\epsilon'$ known from \cite{Leutwyler:96b}
\begin{eqnarray}
\epsilon_0&=&\bar\epsilon_0\cos\theta_0 \frac{\cos\theta_0-\sqrt 2\sin\theta_0}{\cos\theta_0+\sin\theta_0/\sqrt 2}, \nonumber \\
\epsilon_0'&=&\bar\epsilon_0\sin\theta_0 \frac{\sin\theta_0+\sqrt 2\cos\theta_0}{\sin\theta_0-\cos\theta_0/\sqrt 2},
\end{eqnarray}
where the angle $\bar\epsilon_0$ has been obtained by Gross, Treiman, and Wilczek \cite{Gross:79} disregarding the $\eta - \eta'$ mixing   
\begin{equation}
\label{GTW}
\bar\epsilon_0=\frac{\sqrt 3}{4}\frac{m_d-m_u}{m_s-\hat m}=0.011,
\end{equation}
we observe the known effect: The mixing with the $\eta'$ increases significantly the value of the angle $\epsilon_0$ compared to $\bar\epsilon_0$. The LO estimate $\epsilon_0=0.018$ we found is consistent with the estimates $\epsilon\simeq 2\bar\epsilon_0$, for $\theta_0\simeq -22^\circ$ made in \cite{Leutwyler:96b}, and $\epsilon =0.017\pm 0.002$ in \cite{Kroll:05}, both obtained under the same assumptions: The use of Daschen's theorem and $\eta-\eta'$ mixing. This is frustrating because $\epsilon$ enters the amplitude $\eta\to 3\pi$, making the width unacceptably large. This effect was considered in \cite{Leutwyler:96b}, where it was indicated that the problem lies in the accuracy of the LO result. Deviations of order $20-30$\% are to be expected and this does not indicate that the $1/N_c$ expansion fails. It was claimed that the effect can be resolved by taking into account the higher order corrections. Our calculations show that this is exactly what happens. The $1/N_c$ correction $\Delta\epsilon$ leads to complete agreement of our result $\epsilon =0.011$, both with the result of the current algebra and with the result of $\chi$PT. From this we conclude that the LO effect of $\eta$-$\eta'$ mixing on the angle $\epsilon$ is completely offset by the NLO corrections.

Equation (\ref{dN}) must be discussed in a little more detail due to its relation to the low energy constants of $1/N_c\chi$PT given by Eq.\,(\ref{dNcomp}). Recall that $\Delta_N$ is treated as a small parameter, because it represents a term of order $1/N_c$. Indeed, it is reasonably small, our estimate is $\Delta_N =-0.46+0.82=0.36$. It can be seen that the contribution of the gluon anomaly (the second term) differs in sign from the OZI-rule violating contribution (the first term) and dominates. This way the gluon anomaly suppresses the effects of $SU(3)$ and isospin symmetry breaking in $\Delta\mu^2_{08}$ and $\Delta\mu^2_{03}$. Of course, the opposite can also be said: the OZI rule violating interaction (\ref{OZI}) reduces the effect of the gluon anomaly in these channels. 

Further, since the following relations hold
\begin{eqnarray}
&&2\sqrt{6}\frac{\lambda_Z}{F^2}  \leftrightarrow \frac{\Lambda_1}{2}-\Lambda_2, \nonumber \\
&&\lambda_UG_S\frac{a-\delta_M}{2M_0^2} \leftrightarrow 4 L_5 \frac{M_0^2}{F_0^2},
\end{eqnarray}
we obtain the following estimates for the couplings on the right-hand side of these relations, namely, $\Lambda_1/2-\Lambda_2=-0.46$ and $4L_5 M_0^2/F_0^2=0.82$. These values are noticeably lower than the estimates obtained in \cite{Goity:02}, where, for instance, set (NLO No.\,1) gives  the values $-0.65$ and $1.12$ correspondingly. In this case, however, it would be naive to expect complete agreement between the approaches, since the $\Lambda_1$ is also responsible for NLO correction to $\Delta\mu_{00}^2$ in $1/N_c\chi$PT, which, as we have already noted above, is not the case in the $1/N_c$ NJL model. 

\section{Weak-decay coupling constants in the octet-singlet basis}
\label{s5}
To find the decay constants of pseudoscalars we should relate the fields $\phi_a$ ($a=0,8,3$) to the physical eigenstates $P=\eta', \eta, \pi^0$. As we have already learned, a transition to the physical fields $P$ is carried out in two steps 
\begin{equation}
\phi \stackrel{F_{f}}{\longrightarrow}\phi^R\stackrel{U_\theta}{\longrightarrow}P. 
\end{equation}
At the first step, the dimensionless field $\phi =\sum\phi_a\lambda_a$ arising in the effective meson Lagrangian through the exponential parametrization $U=\xi^2 =\exp (i\phi )$ is replaced by the dimensional variable $\phi^R$ given in the same basis (see Eq.\,(\ref{mix})). The symmetric matrix $F_{f}$ (\ref{matrixF}) is worked out in such a way that the kinetic part of the free Lagrangian takes the standard form. Then, at the second step, diagonalizing the mass part of the free Lagrangian by the rotation $U_\theta$ (for definition of matrix $U_\theta$ see Eq.\,(\ref{U})), we come to the physical fields $P=\eta', \eta, \pi^0$. 

The first step of the described procedure generalizes the standard construction of the effective Lagrangian of pseudo Goldstone fields to the case of explicitly broken flavor symmetry. Here \cite{Leutwyler:96b}, the pseudo Goldstone field $\phi$ is also represented by the exponent $U=\exp (i\phi)$, and the pion decay constant $F$ appears in the kinetic part of the effective Lagrangian 
$$
\frac{1}{4}F^2\,\mbox{tr}\left(\partial_\mu U\partial^\mu U^\dagger\right)
$$   
to make the field $\phi$ dimensional by redefining $F\phi=\phi^R$.

In the NJL model the factor at the kinetic part of the Lagrangian arises from a direct calculation of the quark one-loop diagrams. In the case of broken flavor symmetry $m_u\neq m_d\neq m_s$, the place of the factor $F$ is taken by the matrix $F_f$. As a result, to redefine the field $\phi$, it is necessary to use the matrix $F_f$, i.e., $F_f\phi = \phi^R$, and not a simple factor $F$. Obviously, in the chiral limit $F_f$ is a diagonal matrix $F_f= F\,\mbox{diag}(1,1,1)$.

After these general remarks, we find a matrix containing the constants $F_P^{a}$ in their projection onto the octet-singlet basis $a=0,8,3$. This can be achieved by using the product of two transformations $U_\theta F_f$. As a result we have
\begin{equation}
\label{phfields}
P=\sum_{a=0,8,3}\!\! F^a_P \phi_a\!\!=\!\!\left( 
\begin{array}{ccc}
\! F^0_{\eta'} & F^8_{\eta'}  & F^3_{\eta'} \! \\
\! F^0_{\eta}  & F^8_{\eta}  & F^3_{\eta} \! \\
\! F^0_{\pi^0}  & F^8_{\pi^0} & F^3_{\pi^0} \!\\
\end{array}
\right)\!\!
\left( 
\begin{array}{c}
\!\phi_0\! \\ \! \phi_8\! \\ \!\phi_3\! \\
\end{array}
\right),
\end{equation}
where $[F^a_P]=M$, and 
\begin{eqnarray}
\label{constFaP}
&&F^0_{\eta'}\!=\!F_{00} \cos\theta\! +\! F_{08}\sin\theta, \nonumber \\
&&F^8_{\eta'}\!=\!F_{80} \cos\theta\! +\! F_{88}\sin\theta, \nonumber \\
&&F^3_{\eta'}\!=\!F_{30} \cos\theta\! +\! F_{38}\sin\theta\! -\!\epsilon' F_{33},  \nonumber \\
&&F^0_{\eta}\!=\!-F_{00} \sin\theta\! +\! F_{08}\cos\theta, \\
&&F^8_{\eta}\!=\!-F_{80} \sin\theta\! +\! F_{88}\cos\theta, \nonumber  \\
&&F^3_{\eta}\!=\!-F_{30} \sin\theta\! +\! F_{38}\cos\theta \!-\! \epsilon F_{33}, \nonumber \\
&&F^0_{\pi^0}\!=\!F_{03}\!+\!F_{00}(\epsilon'\!\cos\theta\!-\!\epsilon\sin\theta )\!+\! F_{08}(\epsilon'\! \sin\theta\!+\!\epsilon\cos\theta ), \nonumber \\
&&F^8_{\pi^0}\!=\!F_{83}\!+\!F_{80}(\epsilon'\!\cos\theta\!-\!\epsilon\sin\theta )\!+\! F_{88}(\epsilon'\! \sin\theta\!+\!\epsilon\cos\theta ), \nonumber \\
&&F^3_{\pi^0}\!=\!F_{33}. \nonumber 
\end{eqnarray}
Some useful relations between these constants are collected in Appendix \ref{app3}.

To lowest order in $1/N_c$, we have $F_{08}=F_{03}=F_{38}=0$ and then in (\ref{phfields}) we arrive to the standard pattern with one mixing angle $\theta$. 

In the formulas above, it is necessary to take into account only the terms that do not exceed the accuracy of our calculations here. Hence, expanding in powers of $1/N_c$ and retaining only the first two terms, we find   
\begin{eqnarray}
\label{FaPsol}
&&F^0_{\eta'}=f_0\cos\theta_0-F\Delta\theta_{+}\sin\theta_0,   \nonumber\\
&&F^0_{\eta}=-f_0\sin\theta_0 -F\Delta\theta_{+} \cos\theta_0,  \nonumber\\ 
&&F^8_{\eta'}=f_8\sin\theta_0 +F\Delta\theta_{-}\cos\theta_0, \nonumber \\
&&F^8_{\eta}=f_8\cos\theta_0 -F\Delta\theta_{-}\sin\theta_0, \nonumber \\
&&F^3_{\eta'}=-\frac{f_{K^0}\!-\!f_{K^\pm}}{\sqrt 3}\!\left(\sin\theta_0\!+\!\sqrt 2\cos\theta_0\right)\! -\epsilon'_0f_\pi\!-\!\Delta\epsilon' F, \nonumber \\
&&F^3_{\eta}=-\frac{f_{K^0}\!-\!f_{K^\pm}}{\sqrt 3}\!\left(\cos\theta_0\!-\!\sqrt 2\sin\theta_0\right)\! -\epsilon_0f_\pi\!-\!\Delta\epsilon F, \nonumber \\
&&F^0_{\pi^0}=-\sqrt\frac{2}{3}\left(f_{K^0}\!-\!f_{K^\pm}\right)+f_0(\epsilon_0'\cos\theta_0\!-\!\epsilon_0\sin\theta_0) \nonumber\\
&&\ \ \  -F[\Delta\theta_+(\epsilon_0'\sin\theta_0\!+\!\epsilon_0\cos\theta_0)\!-\!\Delta\epsilon'\cos\theta_0\!+\!\Delta\epsilon\sin\theta_0], \nonumber \\
&&F^8_{\pi^0}=-\sqrt\frac{1}{3}\left(f_{K^0}\!-\!f_{K^\pm}\right)+f_8(\epsilon_0'\sin\theta_0\!+\!\epsilon_0\cos\theta_0) \nonumber\\
&&\ \ \  +F[ \Delta\theta_{-}(\epsilon_0'\cos\theta_0\!-\!\epsilon_0\sin\theta_0)\!+\!\Delta\epsilon'\sin\theta_0\!+\!\Delta\epsilon\cos\theta_0], \nonumber \\
&&F^3_{\pi^0}=f_\pi,
\end{eqnarray}
where
\begin{equation}
\Delta\theta_{\pm}\equiv\Delta\theta  \pm \frac{2\sqrt{2}}{3F}(f_K-f_\pi).
\end{equation}

To obtain the numerical values of weak decay constants we use parameter set (b) given in Table \ref{table2}. As a result, we find
\begin{equation}
\label{FaP1}
F^a_P = F \left( 
\begin{array}{ccc}
\! 1.16 & -0.54  & -0.0054 \! \\
\! 0.12 & 1.20  & -0.016 \! \\
\! -0.001 & 0.011 & 1.02 \!\\
\end{array}
\right).
\end{equation} 
The numerical estimations show that the $\eta'$-meson contains a noticeable ($\sim 50$\%) admixture of the octet component $\phi_8$. On the contrary, the $\eta$ meson is nearly a pure octet: The admixture of $\phi_0$ is an order of magnitude lower than $\phi_8$. Note, that the analysis done in the work \cite{Leutwyler:98} led to the same conclusion. The neutral pion is a pure $\phi_3$-state, the admixture of which in the $\eta$ meson is three times greater than in the $\eta'$ state. 

We also get the following estimates for ratios
\begin{eqnarray}
&&\frac{f_8}{f_\pi}=1+\frac{4}{3F}\left(f_K-f_\pi\right)=1.28, \\
&&\frac{f_0}{f_\pi}=1+\frac{2}{3F}\left(f_K-f_\pi\right)=1.14, 
\end{eqnarray}
which perfectly agrees with the values $f_8 = 1.27(2)f_\pi$ and $f_0=1.14(5)f_\pi$ obtained exclusively on the transition form factors of $\eta$ and $\eta'$, reanalyzed in view of the BESIII observation of the Dalitz decay $\eta'\to \gamma e^+ e^-$ in both space- and time-like regions \cite{Escribano:16}. 

The $1/N_c$-corrections to the leading order result allows one to distinguish two mixing angles $\vartheta_8$ and $\vartheta_0$, which are often used in the phenomenological analysis of $\eta$-$\eta'$ data \cite{Feldmann:98}. Indeed, from  Eq.\,(\ref{FaPsol}) one infers
\begin{eqnarray}
\label{2angles}
&&\!\!\!\!\!\!\!\!\!\! F^8_{\eta'}= f_8 \left( \sin\theta_0 +\frac{F}{f_8}\Delta\theta_-\cos\theta_0\right) \nonumber \\
&&=f_8\sin (\theta_0+F\Delta\theta_-/f_8) \equiv f_8 \sin\vartheta_8, \nonumber \\
&&\!\!\!\!\!\!\!\!\!\! F^0_{\eta}= -f_0 \left( \sin\theta_0 +\frac{F}{f_0}\Delta\theta_+\cos\theta_0\right) \nonumber \\
&&=-f_0\sin (\theta_0+F\Delta\theta_+/f_0) \equiv -f_0\sin\vartheta_0. 
\end{eqnarray}
That gives $\vartheta_8=\theta_0+F\Delta\theta_-/f_8=-24.2^\circ$, and
$\vartheta_0=\theta_0+F\Delta\theta_+/f_0=-6.0^\circ$.
Further, if we restrict ourselves only to the first correction, we get
\begin{eqnarray}
&&\vartheta_8=\theta_0+\Delta\theta_- =-26.9^\circ, \nonumber\\
&&\vartheta_0=\theta_0+\Delta\theta_+=-4.6^\circ, \nonumber \\
&&\vartheta_0-\vartheta_8=\frac{4\sqrt 2}{3F}\left(f_K-f_\pi \right).
\end{eqnarray}
This result agrees with a low energy theorem \cite{Leutwyler:98}, which states that the difference between the two angles $\vartheta_0-\vartheta_8$ is determined by $f_K -f_\pi$. The numerical values of the angles again can be compared with the result of \cite{Escribano:16}: $\vartheta_8=-21.2(1.9)^\circ$ and $\vartheta_0=-6.9(2.4)^\circ$. 

\section{Physical content of $\phi_a$}
\label{s6}
Let us establish a connection between the octet-singlet components $\phi_0$, $\phi_8$, and $\phi_3$ and the physical eigenstates $P=\eta', \eta, \pi^0$. For that one needs to know the matrix $\mathcal F^P_a$ in the inverse to the (\ref{phfields}) relation 
\begin{equation}
\phi_a=\!\!\!\! \sum_{P=\eta', \eta, \pi^0}\!\!\! \mathcal F^P_a P \!=\!\left( 
\begin{array}{ccc}
\! \mathcal F_0^{\eta'} & \mathcal  F^{\eta}_0  &\mathcal  F^{\pi^0}_0 \! \\
\! \mathcal  F_8^{\eta'}  & \mathcal  F_8^{\eta}  &\mathcal  F_8^{\pi^0} \! \\
\! \mathcal  F_3^{\eta'}  & \mathcal F_3^{\eta} & \mathcal F_3^{\pi^0} \!\\
\end{array}
\right)\!\!
\left( 
\begin{array}{c}
\!\eta'\! \\ \! \eta\! \\ \!\pi^0\! \\
\end{array}
\right),
\end{equation}
Its dimension is $[\mathcal F^P_a]=M^{-1}$. The entries of the matrix $\mathcal F^P_a=F_{1/f}U_\theta^{-1}$ can be found with the use of formulas (\ref{mixphi}) and (\ref{ortho}).

It is not difficult to establish that the matrix elements $\mathcal F^P_a$ have the form (\ref{constFaP}), where one should substitute $F^a_P\to \mathcal F^P_a$, and $F_f \to F_{1/f}$. Then, expanding the result in the $1/N_c$ series, we find to first nonleading order     
\begin{eqnarray}
&&\mathcal F^{\eta'}_0=\frac{1}{f_0}\cos\theta_0 -\frac{\Delta\theta_{-}}{F}\sin\theta_0, \nonumber \\
&&\mathcal F^{\eta}_0=-\frac{1}{f_0}\sin\theta_0 -\frac{\Delta\theta_{-}}{F}\cos\theta_0, \nonumber \\
&&\mathcal F^{\eta'}_8=\frac{1}{f_8}\sin\theta_0 +\frac{\Delta\theta_{+}}{F}\cos\theta_0, \nonumber \\
&&\mathcal F^{\eta}_8=\frac{1}{f_8}\cos\theta_0 -\frac{\Delta\theta_{+}}{F}\sin\theta_0, \nonumber \\
&&\mathcal F^{\eta'}_3=\frac{f_{K^0}\!-\!f_{K^\pm}}{\sqrt{3}F^2}\left(\sqrt{2}\cos\theta_0\!+\!\sin\theta_0\right) \!-\!\frac{\epsilon_0'}{f_3}\!-\!\frac{\Delta\epsilon'}{F}, \nonumber \\
&&\mathcal F^{\eta}_3=\frac{f_{K^0}\!-\!f_{K^\pm}}{\sqrt{3}F^2}\left(\cos\theta_0\!-\!\sqrt{2}\sin\theta_0\right)\! -\!\frac{\epsilon_0}{f_3}\!-\!\frac{\Delta\epsilon}{F}, \nonumber \\
&&\mathcal F^{\pi^0}_0=\sqrt{\frac{2}{3}}\frac{f_{K^0}\!-\!f_{K^\pm}}{F^2}\!+\!\frac{1}{f_0}\left(\epsilon_0'\cos\theta_0\!-\!\epsilon_0\sin\theta_0\right) \nonumber \\
&&\ \ \ -\frac{\Delta\theta_{-}}{F}\left(\epsilon_0'\sin\theta_0\!+\!\epsilon_0\cos\theta_0\right)\! +\!\frac{\Delta\epsilon'}{F}\cos\theta_0\!-\!\frac{\Delta\epsilon}{F}\sin\theta_0, \nonumber\\
&&\mathcal F^{\pi^0}_8=\frac{f_{K^0}\!-\!f_{K^\pm}}{\sqrt{3}F^2}\!+\!\frac{1}{f_8}\left(\epsilon_0'\sin\theta_0\!+\!\epsilon_0\cos\theta_0\right) \nonumber \\
&&\ \ \ +\frac{\Delta\theta_{+}}{F}\left(\epsilon_0'\cos\theta_0\!-\!\epsilon_0\sin\theta_0\right)\! +\!\frac{\Delta\epsilon'}{F}\sin\theta_0\!+\!\frac{\Delta\epsilon}{F}\cos\theta_0, \nonumber\\
&&\mathcal F^{\pi^0}_3=f_\pi^{-1},
\end{eqnarray}
where 
\begin{eqnarray}
\frac{1}{f_0}&=&\frac{1}{F}\left(2-\frac{f_\pi+2f_K}{3F}\right), \nonumber \\
\frac{1}{f_8}&=&\frac{1}{F}\left(2-\frac{4f_K-f_\pi}{3F}\right), \nonumber \\
\frac{1}{f_3}&=&\frac{1}{F}\left(2-\frac{f_\pi}{F}\right).
\end{eqnarray}

The elements of matrix $\mathcal F_a^P$ numerically are  
\begin{equation}
\label{FaP2}
\mathcal F_a^P = \frac{1}{F} \left( 
\begin{array}{ccc}
\! 0.76 & 0.42  & 0.011 \! \\
\! -0.0074 & 0.73  & 0.01 \! \\
\! 0.0012 & -0.0071 & 0.98 \!\\
\end{array}
\right).
\end{equation} 
Thus, we see that the singlet component, in addition to the leading contribution of $\eta'$, has a noticeable admixture of $\eta$. The latter dominates in the octet component $\phi_8$, while the neutral pion dominates in $\phi_3$.

Since matrices $F_P^a$ and $\mathcal F^P_a$ are mutually inverse, one would expect that the numerical result (\ref{FaP2}) should be reasonably close (up to the approximations made) to the result obtained by the direct inversion of matrix (\ref{FaP1}). This is indeed the case for all elements of matrix (\ref{FaP2}) except $\mathcal F_8^{\eta'}$. The latter turns out to be an order of magnitude smaller than such a qualitative estimate gives: $\mathcal F_8^{\eta'}=-0.082/F$. Here there is a significant compensation between two terms: $\mathcal F_8^{\eta'}=(0.175-0.182)/F$. The first term is the $SU(3)$ breaking contribution with cosine, the second one is $(F/f_8)\sin\theta_0$. In the limit of exact $SU(3)$ symmetry this term gives $-0.258$. The $SU(3)$-breaking correction in $f_8$ makes it to be equal $-0.182$. This combined effect of mixing angle $\theta_0$ and $SU(3)$ breaking pushes $\eta'$ out of the octet.    

\section{The strange-nonstrange mixing scheme}
\label{s7}
Instead of a flavor octet-singlet basis, one could choose a scheme with a mixture of strange and non-strange components. It is also often used in the literature (the so-called Feldmann-Kroll-Stech scheme \cite{Feldmann:98}). Therefore, we will briefly focus on it here. To do this, just as we did in Appendix\,\ref{app1}, let us represent the field $\phi^R$ by its components in the orthogonal basis $\lambda_\alpha=(\lambda_S, \lambda_q, \lambda_3)$
\begin{equation}
\phi^R =\!\!\!\!\sum_{\alpha=S,q,3}\!\!\phi_\alpha^R \lambda_\alpha,
\end{equation}
where $\mbox{tr}(\lambda_\alpha\lambda_\beta )=2\delta_{\alpha\beta}$, and 
\begin{eqnarray}
&&\lambda_q=\frac{1}{\sqrt{3}}\left(\sqrt 2\lambda_0+\lambda_8 \right), \nonumber \\
&&\lambda_S=\sqrt 2\lambda_s=\frac{1}{\sqrt{3}}\left(\lambda_0-\sqrt 2\lambda_8 \right).
\end{eqnarray}

It follows then 
\begin{equation}
\phi^R_\alpha = \frac{1}{2}\!\! \sum_{a=0,8,3}\!\!\!\phi^R_a\, \mbox{tr} (\lambda_a\lambda_\alpha)
= \!\!\!\! \sum_{a=0,8,3}\!\!\! U^{-1}_{\alpha a}(\theta_{id},0,0) \phi^R_a,
\end{equation}
or 
\begin{equation}
\label{qsbasis}
\tilde \phi^R \equiv \left( 
\begin{array}{c}
\!\phi_S^R\! \\ \! \phi_q^R \! \\ \!\phi^R_3\! \\
\end{array}
\right) = \left(
\begin{array}{ccc}
\! \cos\theta_{id} & -\sin{\theta_{id}}  & 0 \! \\
\! \sin{\theta_{id}}  &  \cos\theta_{id} & 0 \! \\
\! 0  & 0 & 1 \!  \\
\end{array}
\right) \!\! 
\left( 
\begin{array}{c}
\!\phi^R_0\! \\ \! \phi^R_8\! \\ \!\phi^R_3\! \\
\end{array}
\right),
\end{equation}
where the ideal mixing angle $\theta_{id}=\arctan\sqrt 2\simeq 54.7^\circ$.

Now, it is easy to establish from (\ref{ortho}) that the physical states $P=(\eta',\eta,\pi^0)$ are the linear combinations of states $\tilde\phi^R=(\phi^R_S,\phi_q^R,\phi^R_3)$, namely
\begin{equation}
P=U(\theta,\epsilon,\epsilon') U(\theta_{id},0,0) \tilde \phi^R=U(\varphi,\epsilon,\epsilon') \tilde\phi^R,
\end{equation} 
where the angle $\varphi=\theta +\theta_{id}$, or numerically $\varphi = 39.0^\circ$, and the matrix $U$ is defined in (\ref{U}). This result agrees with  phenomenological estimates $\varphi=39.3^\circ\pm 1.0^\circ$ in \cite{Feldmann:98}, and lattice data obtained by ETM Collaboration \cite{ETM:18} $\varphi=38.8^\circ\pm 2.2^\circ \pm 2.4^\circ$, where the second errors refer to uncertainties induced by chiral extrapolations to the physical point. The anomaly sum rule approach to the transition form factors with a systematic account of the $\eta$-$\eta'$ mixing and quark-hadron duality gives a bit smaller result: $\varphi = (38.1\pm 0.5)^\circ$ \cite{Teryaev:13}.

\section{Dependence on the regularization scheme} 
\label{s8}
The results presented above are based on a well-defined regularization scheme, proper-time regularization. In this case, two integrals with quadratic $J_0(M_i)$ and logarithmic $J_1(M_i)$ divergences are regulated by subtracting off suitable counterterms
\begin{equation}
J_\alpha (M_i)=\int_0^\infty \frac{dt}{t^{2-\alpha}}\rho_{t,\Lambda}e^{-tM_i^2}
\end{equation} 
at the scale $\Lambda$, $\rho_{t,\Lambda}=1-(1+t\Lambda^2)e^{-t\Lambda^2}$. The choice of the regularizing kernel is made in such a way that the expressions for regularized integrals coincide with the one obtained in the covariant four-dimensional Euclidean regularization scheme. This implies, in particular, the following form of the gap equation 
\begin{equation}
\label{gap}
f(M_i,\Lambda )\equiv M_i\left(1-\frac{N_cG_S}{2\pi^2}J_0(M_i) \right)=m_i.
\end{equation}  

 Let us recall that the choice of regularization essentially determines the NJL model. A reasonable regularization must satisfy two main criteria: the minimum effective potential condition should lead to a gap equation, and the vacuum corresponding to the nontrivial solution should possess the Goldstone modes. It is also desirable that the regularization does not break the symmetry of the model. All these requirements are met by the Fock-Schwinger method used here. It is one of the standard regularization prescriptions often used in the literature \cite{Klevansky:92,Cvetic:97}. 

\begin{figure}
    \centering
    \begin{minipage}{0.45\textwidth}
        \centering
        \includegraphics[width=0.9\textwidth]{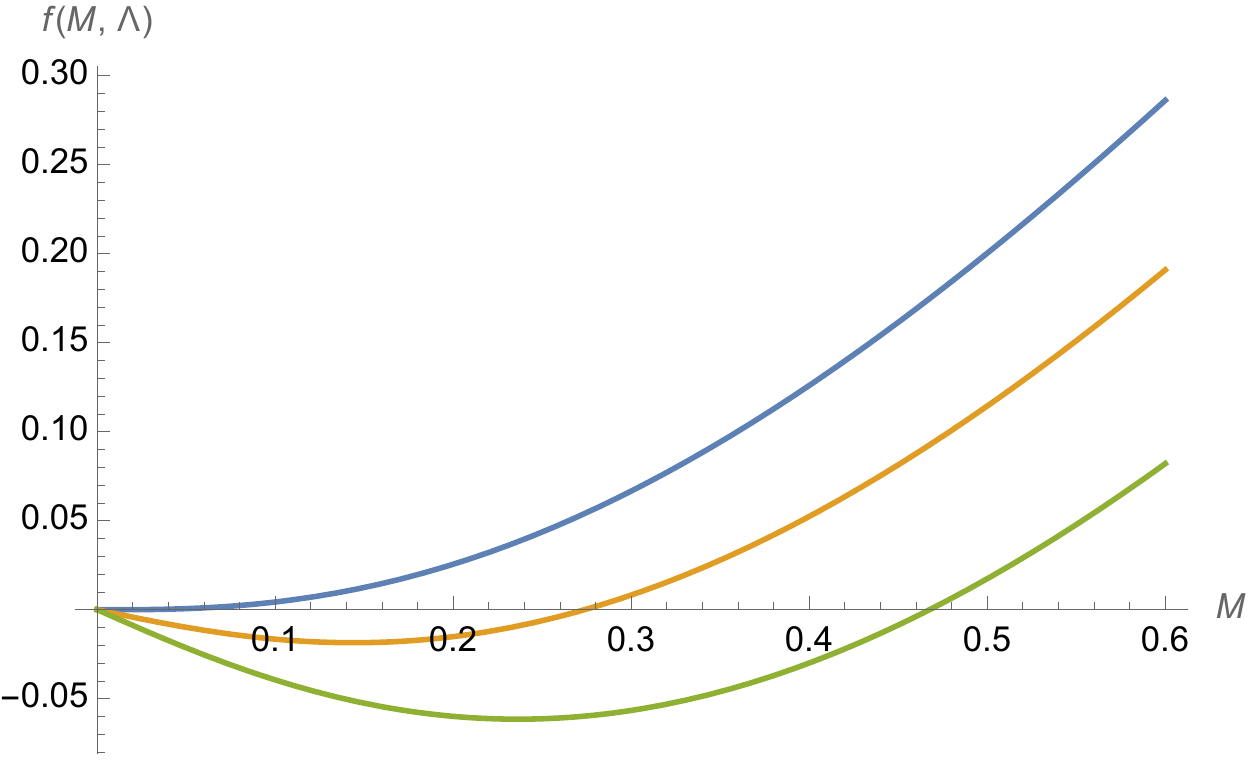} 
        \caption{The left side of the equation (\ref{gap}) as a function of $M_i$ (both in GeV) is shown for the three cutoff values: $\Lambda=1.0\,\mbox{GeV}$, $\Lambda = 1.1\,\mbox{GeV}$ and $\Lambda = 1.2\,\mbox{GeV}$ (top down).}
        \label{fig1} 
    \end{minipage}\hfill
    \begin{minipage}{0.45\textwidth}
        \centering
        \includegraphics[width=0.9\textwidth]{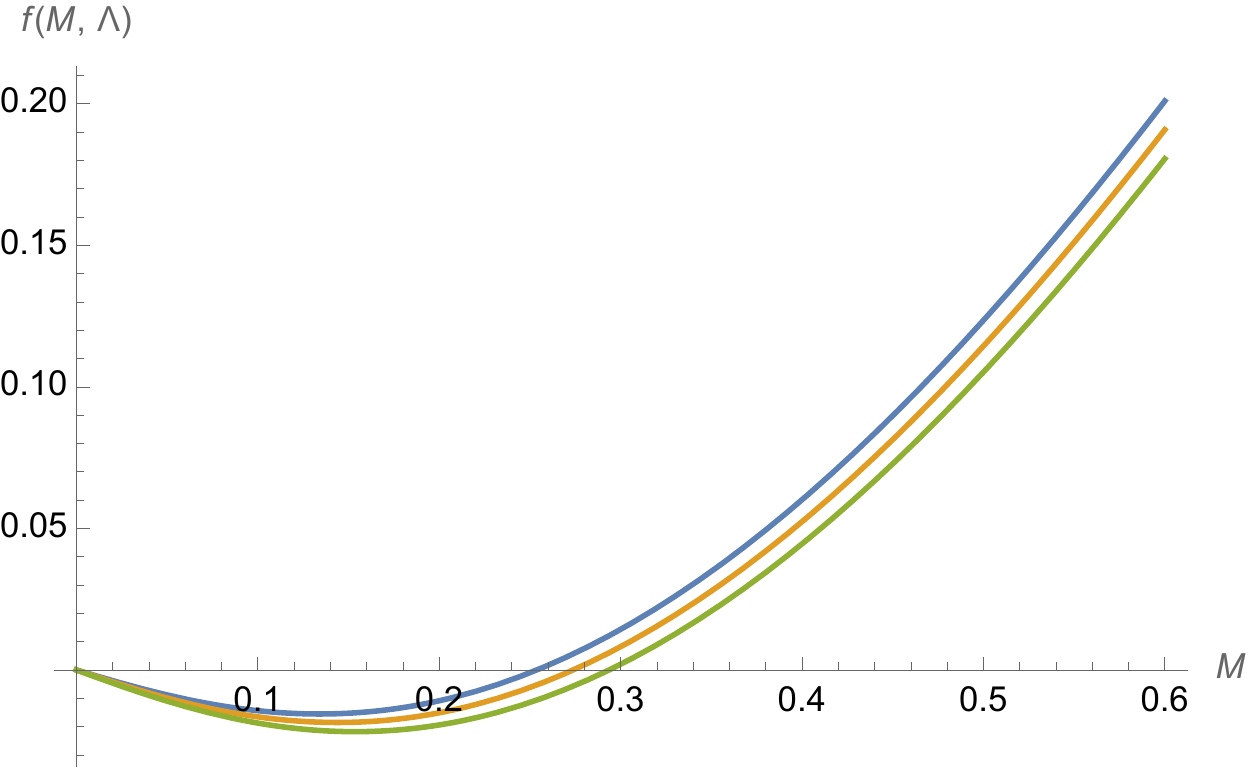} 
        \caption{The same as in Fig.1, but with the $1\%$ deviation in values of $\Lambda$, namely $\Lambda=1.09\,\mbox{GeV}$, $\Lambda = 1.10\,\mbox{GeV}$ and $\Lambda = 1.11\,\mbox{GeV}$ (top down).}
        \label{fig2} 
    \end{minipage}
\end{figure}

The gap equation for $m_i=0$ and a fixed value of $G_S$, 
\begin{equation}
 G_S\Lambda^2 > \frac{2\pi^2}{N_c}=6.58,
\end{equation}
has a solution $M_0(\Lambda)$ that changes drastically with $\Lambda$. Fig.\ref{fig1} shows three curves corresponding to the values of $\Lambda$ deviating from the value chosen in our work $\Lambda =1.1\,\mbox{GeV}$ by 10\%. Such behavior is associated with the original quadratic divergence of the integral $J_0(M_0)$, and is typical for any of the regularization schemes commonly used in the NJL model. In particular, one can obtain that $M_0 (1.0\,\mbox{GeV}) =19.7\, \mbox{MeV}$, $M_0 (1.1\,\mbox{GeV}) =274\, \mbox{MeV}$, $M_0 (1.2\,\mbox{GeV}) =468\, \mbox{MeV}$ for $G_S=6.6\,\mbox{GeV}^{-2}$. It shows that $M_0$ changes a lot with $\Lambda$, and it does make sense to consider only $1\%$ (or less) deviations in the value of $\Lambda$. This case is shown in Fig.\ref{fig2}. It can be seen that the corresponding solutions of the gap equation differ only marginally 
\begin{equation}
\Lambda = 1.10\pm 0.01\,\mbox{GeV}, \quad M_0=274\pm 20\,\mbox{MeV},
\end{equation}
i.e., a $1\%$ change in the value of the cutoff $\Lambda$ leads to about $7\%$ changes in the value of $M_0$. This can be a source of theoretical uncertainties associated with the regularization scheme used.

Since all other vacuum characteristics are expressed in terms of these two parameters (and $G_V=7.4\,\mbox{GeV}^{-2}$), it is easy to establish that 
\begin{eqnarray}
a&=&3.50^{-0.28}_{+0.33}, \quad \delta_M=0.67^{+0.10}_{-0.12}, \nonumber \\
F&=&90.54^{+2.82}_{-3.10}\,\mbox{MeV}, \quad |\langle \bar qq\rangle_0^{1/3}|=275^{+6}_{-7} \,\mbox{MeV}.
\end{eqnarray}
Note that a $1\%$ change in the value of $\Lambda$ leads to a $3\%$ change in the value of the constant $F$, i.e., the physical constant $f_\pi$ coincides with $F$ within the error. Indeed, our estimates give 
\begin{equation}
f_\pi=92.22^{+2.54}_{-2.74}.
\end{equation} 
It should be noted that current knowledge about the constant $f_\pi=92.277(95)\,\mbox{MeV}$ \cite{PDG} imposes such a strict boundaries on the range of values admissible for the cutoff $\Lambda$ (or alternatively on the coupling of four-quark interactions $G_S$), that one could reach such accuracy in the NJL model only by the precise tuning of $\Lambda=1100.00^{+0.55}_{-0.16}\,\mbox{MeV}$.  

It is also interesting to check how successful the NJL model predictions are for the low-energy constants $L_5$ and $L_8$ of the $1/N_c\chi$PT which are scale independent at the order considered. Here the $1/N_c$ NJL model gives
\begin{eqnarray}
\label{L58}
L_5&=&\frac{a-\delta_M}{8M_0^2}G_SF^4=\left(2.1^{-0.3}_{+0.4}\right)10^{-3}, \nonumber \\
L_8&=&\frac{a}{16M_0^2}G_SF^4=\left(1.3\mp 0.1\right)10^{-3}.  
\end{eqnarray}  
The values of these constants are consistent with their phenomenological estimates made in $\chi$PT: $L_5=(2.2\pm0.5)10^{-3}$, $L_8=(1.1\pm0.3)10^{-3}$ \cite{Gasser:85}, however these values have to be redetermined using the logic of the mixed expansion scheme adopted in the $1/N_c\chi$PT. Such calculations were made, for example, in \cite{Goity:02}. Neglecting the contributions of chiral logarithms, the authors obtained the following estimates: $2L_5+L_8=(5.26\pm 0.01)10^{-3}$, $2L_8-L_5=(0.8\pm 0.9)10^{-5}$. Our result (\ref{L58}) for these combinations leads to the values $2L_5+L_8=(5.47^{-0.73}_{+0.89})10^{-3}$, $2L_8-L_5=(0.50^{+0.06}_{-0.09})10^{-5}$. The NLO analysis of the $\eta$-$\eta'$ mixing in \cite{Bickert:17} gives $L_5=(1.86\pm 0.06)10^{-3}$, and $L_8=(0.78\pm 0.05)10^{-3}$. As pointed out in \cite{Kaplan:86}, $L_8$ cannot be determined on purely phenomenological grounds. As a consequence, the sign of the difference $2L_8-L_5$ is fixed only in the framework of a specific model. We see that in the $1/N_c$ NJL model the sign is positive
\begin{equation}
\label{sign}
2L_8-L_5=\frac{\delta_M}{8M_0^2}G_SF^4>0,
\end{equation}
in contrast to the estimates in \cite{Bickert:17}. It is clear that inequality (\ref{sign}) is ensured by a positive value of $\delta_M=0.67$ (with a $15\%$ theoretical uncertainty).

This analysis can be extended to all the estimates we obtained above. This will be done a little later, when we have accumulated enough theoretical material on various physical properties and processes involving pseudoscalar mesons.  

\section{Conclusions} 
\label{s9}
We continued studying the properties of the nonet of pseudoscalar mesons in the NJL model, where we changed the counting rule for the masses of current quarks. In the recent work \cite{Osipov:23} the charged states were considered, and in this work the same tool has been applied to calculate the main characteristics of the neutral members of the nonet. 

The realistic description of neutral modes is impossible without the use of Lagrangians responsible for breaking the axial $U(1)_A$ symmetry and OZI rule. They are well known, so we did not set ourselves the goal of obtaining them on the basis of corresponding multi-quark interactions. 

Four-quark interactions generate the kinetic part of the free Lagrangian and make a major contribution to the particle masses. We calculated these one-loop diagrams and demonstrated that the singlet component $\phi_0$ (as well as other two $\phi_3$ and $\phi_8$) is a superposition of three eigenvectors $\phi_a^R$, $a=0,3,8$ which diagonalize the kinetic part of free Lagrangian in the octet-singlet basis. We have shown that this is a direct result of explicit symmetry breaking in the superconducting model.       

Since the singlet field $\phi_0$ acquires a mass term due to the gluon anomaly, the following transition to the eigenstates $\phi_a^R$ induces contributions to the off-diagonal elements $\Delta\mu_{08}^2$ and $\Delta\mu_{03}^2$ of the mass matrix already in the next order of the $1/N_c$ expansion. We have shown that due to this mechanism, the gluon anomaly substantially weakens the effects of explicit chiral symmetry breaking observed in the leading order.

It is interesting to note that this mechanism copes with the problem that Leutwyler pointed out at the time: The value of the $\pi^0$-$\eta$ mixing angle $\epsilon$ turns out to be too large if we restrict ourselves to only the LO contribution. This will have a catastrophic effect on the width of the $\eta\to 3\pi$ decay. We have shown that taking into account the NLO correction eliminates this difficulty.

The mass matrix of neutral states is diagonalized by a rotation parameterized by three mixing angles $\theta$, $\epsilon$ and $\epsilon'$. In particular, rotation through the angle $\theta$ eliminates the $\eta$-$\eta'$ mixing. However, when considering the decay constants of $\eta$ and $\eta'$ mesons, one angle $\theta$ is not enough. Therefore, a picture with two mixing angles is often used. We have shown that NLO corrections effectively lead to a two-angle pattern, but unfortunately could not give its full theoretical justification.

It should be emphasized that the above results would be more complete if calculations of electromagnetic decays of pseudoscalar mesons were carried out. We are currently working on this issue.

\section*{Acknowledgments}
I am grateful to B. Hiller for the interest in the subject and useful discussions. I am also grateful to H. Weigel for helpful discussions of some details of the two-mixing-angle scheme. This work is supported by Grant from Funda\c{c}\~ ao para a Ci\^ encia e Tecnologia (FCT) through the Grant No. CERN/FIS-COM/0035/2019. 

\appendix
\section{Rescaling of neutral fields }
\label{app1}
In this appendix we obtain some useful relations between the neutral fields before and after rescaling. 

Recall that the neutral field $\phi$ can be represented by its components taking values in the algebra of the $U(3)$ group or more specifically in the subset of the diagonal hermitian generators $\lambda_a$ or their linear combination
\begin{equation}
\label{Phi-field}
\phi=\!\!\!\!\sum_{a=0,8,3}\!\!\phi_a\lambda_a=\!\!\!\!\sum_{i=u,d,s}\!\!\phi_i\lambda_i\end{equation}  
In what follows, we will use either a octet-singlet basis $(\lambda_0, \lambda_8, \lambda_3)$, or the flavor one $(\lambda_u, \lambda_d, \lambda_s)$ which are related in a standard way  
\begin{eqnarray}
\lambda_u&=&\frac{\lambda_3}{2}+\frac{\sqrt 2\lambda_0+\lambda_8}{2\sqrt 3}=\mbox{diag}(1,0,0),
\nonumber \\
\lambda_d&=&-\frac{\lambda_3}{2}+\frac{\sqrt 2\lambda_0+\lambda_8}{2\sqrt 3}=\mbox{diag}(0,1,0), \nonumber \\
\lambda_s&=&\frac{\lambda_0-\sqrt 2 \lambda_8}{\sqrt 6}=\mbox{diag}(0,0,1).
\end{eqnarray}
Obviously that $\mbox{tr}\,\phi^2\!=\!2\sum\phi_a^2\!=\!\sum\phi_i^2$. As a consequence we also have
\begin{equation}
\label{Fields}
\phi_i=\!\!\! \sum_{a=0,8,3}\!\!\! O_{ia}\phi_a, \quad \phi_a=\!\!\!\sum_{i=u,d,s}\!\!\! O^{-1}_{ai}\phi_i, 
\end{equation}
where $\phi_i\!=\!(\phi_u,\phi_d,\phi_s)$, $\phi_a\!=\!(\phi_0,\phi_8,\phi_3)$, and
\begin{equation}
\label{O}
 O=\frac{1}{\sqrt 3}\left( 
\begin{array}{ccc}
\sqrt 2 & 1 & \sqrt 3   \\
\sqrt 2 & 1 & -\sqrt 3 \\
\sqrt 2 & -2 & 0 \\
\end{array}
\right), \quad  O^{-1} =\frac{1}{2} O^T.
\end{equation}
It yields the property $\sum_a O_{ia}O_{ja}=2\delta_{ij}$.

The kinetic part of the free Lagrangian takes the standard form if one rescaled the flavor components 
\begin{equation}
\label{rescaling}
f_i\phi_i=\phi_i^R.
\end{equation} 
The rescaled field $\phi^R$ can also be charactirized by its components in any of $\lambda$-matrix bases
\begin{equation}
\label{Rfields}
\phi^R=\!\!\!\!\sum_{i=u,d,s}\!\!\phi_i^R\lambda_i=\!\!\!\!\sum_{a=0,8,3}\!\!\phi_a^R\lambda_a.
\end{equation}  

Our task here is to find out how rescaling (\ref{rescaling}) modifies the octet-singlet components $\phi_a$. As we will show, in octet-singlet components rescaling (\ref{rescaling}) has a non-diagonal form. Thus, we need to find the matrix $F$ and its inverse one $F^{-1}$ in the relations  
\begin{equation}
\label{mix}
\phi_a^R=\sum_b F_{ab} \phi_b, \quad \phi_a=\sum_b (F^{-1})_{ab} \phi_b^R.
\end{equation} 

Given that $\lambda_i\! =\!\sum_a\! O_{ia}\lambda_a/2$, we find from the left-hand side of Eq.\,(\ref{Rfields}) and Eq.\,(\ref{Fields})
\begin{equation}
\sum_i f_i \phi_i \lambda_i =\sum_{i,b} f_i O_{ib}\phi_b\lambda_i=\frac{1}{2}\sum_{i,b,a} f_i O_{ib} O_{ia}\phi_b\lambda_a.
\end{equation}
Comparing the result with the right-hand side of Eq.\,(\ref{Rfields}) we conclude that
\begin{equation}
\label{matrixF}
F_{ab}=\frac{1}{2}\sum_{i=u,d,s}\!\!\! O_{ia} O_{ib}  f_i.
\end{equation}
It follows then  
\begin{eqnarray}
\label{elementsF}
&& F_{00}=\frac{1}{3}\left(f_u+f_d+f_s\right), \nonumber \\
&& F_{08}=F_{80}=\frac{1}{3\sqrt 2}\left(f_u+f_d-2f_s\right), \nonumber \\
&& F_{03}=F_{30}=\frac{1}{\sqrt 6}\left(f_u-f_d\right), \nonumber \\
&&F_{88}=\frac{1}{6}\left(f_u+f_d+4f_s\right), \nonumber \\
&&F_{38}=F_{83}=\frac{1}{2\sqrt 3}\left(f_u-f_d\right), \nonumber \\
&&F_{33}=\frac{1}{2}\left(f_u+f_d\right). 
\end{eqnarray}

Starting from equation (\ref{Phi-field}) and acting in a similar way, we find
\begin{equation}
\sum_i \frac{\phi_i^R}{f_i} \lambda_i =\sum_{i,b} \lambda_i O_{ib}\frac{\phi^R_b}{f_i}=\frac{1}{2}\sum_{i,b,a} O_{ib} O_{ia} \frac{\phi_b^R}{f_i} \lambda_a.
\end{equation}
That gives
\begin{equation}
\mathcal (F^{-1})_{ab}=\frac{1}{2}\sum_{i=u,d,s}\!\!\! O_{ia} O_{ib}  \frac{1}{f_i}.
\end{equation}
It follows that the elements of the inverse matrix are obtained from the formulas (\ref{elementsF}) by replacing $f_i\to 1/f_i$. Formally, if the notation of matrix $F$ explicitly specifies its dependence on $f_i$, namely $F_f$, then for the inverse matrix $F^{-1}$ we can use the shorthand $F_{1/f}$.    

In particular, the second equation in (\ref{mix}) takes the form 
\begin{eqnarray}
\label{mixphi}
\phi_0 &=& \frac{\phi_0^R}{f_0}\!+\! \left(\frac{1}{f_u}\!-\!\frac{1}{f_d} \right) \frac{\phi_3^R}{\sqrt 6}  
\!+\! \left(\frac{1}{f_u}\!+\!\frac{1}{f_d}\!-\!\frac{2}{f_s}\right) \frac{\phi_8^R}{3\sqrt 2},  \nonumber \\            
\phi_8 &=& \frac{\phi_8^R}{f_8} \!+\! \left(\frac{1}{f_u}\!-\!\frac{1}{f_d} \right) \frac{\phi_3^R}{2\sqrt 3} 
\!+\!\left(\frac{1}{f_u}\!+\!\frac{1}{f_d}\!-\!\frac{2}{f_s}\right) \frac{\phi_0^R}{3\sqrt 2}, \nonumber\\
\phi_3 &=& \frac{\phi_3^R}{f_3} \!+\! \left(\frac{1}{f_u}\!-\!\frac{1}{f_d}\right)
\frac{\phi_8^R \!+\!\sqrt 2 \phi_0^R}{2\sqrt 3},               
\end{eqnarray}
where the following notations are used
\begin{eqnarray}
\label{f083}
f_0^{-1}&=&\frac{1}{3} \left(f_u^{-1} \!+\!f_d^{-1}\! +\!f_s^{-1}\right), \nonumber \\
f_8^{-1}&=&\frac{1}{6} \left(f_u^{-1} \!+\!f_d^{-1}\! +\!4f_s^{-1}\right),  \nonumber \\
f_3^{-1}&=&\frac{1}{2} \left(f_u^{-1}\!+\!f_d^{-1} \right). 
\end{eqnarray}

\section{Diagonalization of the mass matrix and physical states}
\label{app2}
Let us recall some useful details of the diagonalization procedure of the mass matrix (\ref{massmatrix}). For that we use the transformation 
\begin{equation}
\label{RtoPh}
\left( 
\begin{array}{c}
\phi_0^R \\ \phi_8^R \\ \phi_3^R \\
\end{array}
\right)
=U^{-1}(\theta, \epsilon, \epsilon' )
\left( 
\begin{array}{c}
\! \eta' \! \\ \! \eta \! \\ \! \pi^0 \!  \\
\end{array}
\right).
\end{equation}
where $U^{-1}$ is a matrix defined by
\begin{equation}
\label{U}
U^{-1}(\theta, \epsilon, \epsilon' )=\left( 
\begin{array}{ccc}
\! \cos\theta & -\sin\theta & \epsilon'\cos\theta\!-\!\epsilon \sin\theta\!    \\
\! \sin\theta  & \cos\theta & \epsilon' \sin\theta\!+\!\epsilon \cos\theta\! \\
\! -\epsilon' & -\epsilon & 1 \!\\
\end{array}
\right).
\end{equation}
The matrix is an element of $SO(3)$ which is parametrized by three angles $\theta$, $\epsilon$, $\epsilon'$. The first arises from the mass difference of the strange and nonstrange quarks and breaks $SU(3)$, i.e., in the limit of exact $SU(3)$ symmetry $\theta\to 0$. The other two angles describe the isospin breaking effects. They are proportional to the difference $m_d-m_u$. This factor is small, thus we systematically neglect the higher powers of $\epsilon$ and $\epsilon'$.    

The considered orthogonal transformation diagonalizes the mass matrix $\mathcal M^2$ if the mixing angles satisfy the conditions
\begin{eqnarray}
&&\epsilon = \left( \mathcal M^2_{03} \sin\theta - \mathcal M^2_{38} \cos\theta \right)/(m_{\eta}^2-m^2_{\pi^0}), \nonumber \\
&&\epsilon'\!=\! - \left( \mathcal M^2_{03}\cos\theta + \mathcal M^2_{38}\sin\theta \right) / (m_{\eta'}^2-m^2_{\pi^0}), \nonumber \\
&&\tan 2\theta=2\mathcal M^2_{08}/ \left( \mathcal M^2_{00}-\mathcal M^2_{88}\right),  
\end{eqnarray}
where the masses of neutral states are  
\begin{eqnarray}
m_{\eta, \eta'}^2&\!=\!&\frac{1}{2}\!\left(\mathcal M^2_{00}\!+\!\mathcal M^2_{88}\!\mp\! \sqrt{(\mathcal M^2_{00}\!-\!\mathcal M^2_{88})^2\!+\!4\mathcal M^4_{08} }\right)\!, \nonumber\\
m_{\pi^0}^2&\!=\!&\mathcal M^2_{33}.
\end{eqnarray}

Since the mass matrix (\ref{massmatrix}) is the sum of the leading contribution and the first correction to it, then the angles should be sought in a similar form, namely $\theta =\theta_0+\Delta\theta$, $\epsilon = \epsilon_0+\Delta\epsilon$, and $\epsilon' = \epsilon_0'+\Delta\epsilon'$. The angles $\theta_0$, $\epsilon_0$ and $\epsilon_0'$ are of order $N_c^0$. They are responsible for diagonalizing the leading contribution. The extra terms $\Delta\theta$, $\Delta\epsilon$ and $\Delta\epsilon'$ are of order $1/N_c$. They are responsible for diagonalizing the mass matrix with corrections included.
\begin{eqnarray}
&&\tan 2\theta_0=\frac{2\mu^2_{08}}{\mu^2_{00}-\mu^2_{88}}, \nonumber \\
&&\Delta\theta =\frac{1}{4}\sin 4\theta_0\left(\frac{\Delta\mu_{08}^2}{\mu_{08}^2}-\frac{\Delta\mu_{00}^2-\Delta\mu_{88}^2}{\mu_{00}^2-\mu_{88}^2} \right), \nonumber \\
&&\epsilon_0=\left( \mu^2_{03} \sin\theta_0 - \mu^2_{38} \cos\theta_0 \right)/(\mu_{\eta}^2-\mu^2_{\pi^0}), \nonumber \\
&&\Delta\epsilon =\Delta\theta\,\frac{\mu_{03}^2\cos\theta_0+\mu_{38}^2\sin\theta_0}{\mu_\eta^2-\mu_{33}^2} \nonumber \\
&&+\frac{\Delta\mu_{03}^2\sin\theta_0-\Delta\mu_{38}^2\cos\theta_0}{\mu_\eta^2-\mu_{33}^2} \nonumber\\
&&-\frac{\mu_{03}^2\sin\theta_0-\mu_{38}^2\cos\theta_0}{(\mu_\eta^2-\mu_{33}^2)^2}\left(\Delta\mu^2_\eta -\Delta\mu_{33}^2\right), \nonumber \\
&&\epsilon'_0= - \left( \mu^2_{03}\cos\theta_0 + \mu^2_{38}\sin\theta_0 \right) / (\mu_{\eta'}^2-\mu^2_{\pi^0}), \nonumber \\
&&\Delta\epsilon'\!=\!\Delta\theta\,\frac{\mu_{03}^2\sin\theta_0-\mu_{38}^2\cos\theta_0}{\mu_{\eta'}^2-\mu_{33}^2} \nonumber \\
&&-\frac{\Delta\mu_{03}^2\cos\theta_0+\Delta\mu_{38}^2\sin\theta_0}{\mu_{\eta'}^2-\mu_{33}^2} \nonumber\\
&&+\frac{\mu_{03}^2\cos\theta_0+\mu_{38}^2\sin\theta_0}{(\mu_{\eta'}^2-\mu_{33}^2)^2}\left(\Delta\mu^2_{\eta'} -\Delta\mu_{33}^2\right), 
\end{eqnarray}

The eigenvalues are the squares of the $\eta$, $\eta'$, $\pi^0$ masses
\begin{eqnarray}
&&m_{\eta, \eta'}^2=\mu_{\eta, \eta'}^2+\Delta \mu_{\eta, \eta'}^2, \nonumber \\
&&m_{\pi^0}^2=\mu_{33}^2+\Delta \mu_{33}^2, 
\end{eqnarray}
where 
\begin{eqnarray}
&&\mu_{\eta, \eta'}^2\!=\!\frac{1}{2}\!\left[\mu^2_{00}\!+\!\mu^2_{88}\!\mp\! \sqrt{(\mu^2_{00}\!-\!\mu^2_{88})^2\!+\!4\mu^4_{08} }\right]\!, \nonumber\\
&&\Delta\mu_{\eta, \eta'}^2\!=\!\frac{1}{2}\left(\frac{}{}\Delta\mu_{00}^2+\Delta\mu_{88}^2  \right. \nonumber \\
&&\left. \mp \frac{(\mu^2_{00}\!-\!\mu^2_{88})(\Delta\mu^2_{00}\!-\!\Delta\mu^2_{88})+4\mu^2_{08}\Delta\mu^2_{08}}{\sqrt{(\mu^2_{00}\!-\!\mu^2_{88})^2\!+\!4\mu^4_{08} }} \right), \nonumber\\
&&m_{\pi^0}^2=\bar m_{\pi^\pm}^2.
\end{eqnarray}
It is these formulas that are used to fix the model parameters by masses of $\eta$ and $\eta'$ mesons.

\section{Some useful relations}
\label{app3}
Eqs.\,(\ref{constFaP}) imply a number of linear and quadratic relations between coupling constants. 

The linear relations are
\begin{eqnarray}
\label{LRR}
&&F^0_{\eta'} \cos\theta -F^0_{\eta} \sin\theta = F_{00}, \nonumber \\ 
&&F^0_{\eta'} \sin\theta +F^0_{\eta} \cos\theta = F_{08}, \nonumber \\
&&F^8_{\eta'} \cos\theta -F^8_{\eta} \sin\theta = F_{08},  \\
&&F^8_{\eta'} \sin\theta +F^8_{\eta} \cos\theta = F_{88}, \nonumber \\
&&F^3_{\eta'} \cos\theta -F^3_{\eta} \sin\theta = F_{03}+F_{33}(\epsilon \sin\theta -\epsilon'\cos\theta ), \nonumber \\
&&F^3_{\eta'} \sin\theta +F^3_{\eta} \cos\theta = F_{38}-F_{33}(\epsilon \cos\theta +\epsilon'\sin\theta ). \nonumber
\end{eqnarray}

The second order relations are
\begin{eqnarray}
\label{QR}
&&\left(F^8_{\eta} \right)^2 +\left(F^8_{\eta'} \right)^2 =\left(F_{88} \right)^2+\left(F_{08} \right)^2, \nonumber \\
&&\left(F^0_{\eta} \right)^2 +\left(F^0_{\eta'} \right)^2 =\left(F_{00} \right)^2+\left(F_{08} \right)^2, \nonumber \\
&&F^8_{\eta} F^0_{\eta} +F^8_{\eta'} F^0_{\eta'} =F_{88}F_{08} + F_{00}F_{08}.
\end{eqnarray}

From Eqs.\,(\ref{FaPsol}) we obtain the analogue of the Gell-Mann-Okubo formula
\begin{equation}
\label{GOf}
(F_\eta^8)^2+(F_{\eta'}^8)^2=f_8^2 =\frac{1}{3}\left(4f_K^2-f_\pi^2 \right), 
\end{equation}
and other well-known relation
\begin{equation}
\label{R2}
F_\eta^8F_\eta^0+F_{\eta'}^8F_{\eta'}^0=\sqrt 2(f_0^2-f_8^2)=\frac{2\sqrt 2}{3}\left(f_\pi^2-f_K^2\right).
\end{equation}
Both of them are valid to first nonleading order.

If in these formulas we put 
\begin{eqnarray}
\label{F8}
&&F_\eta^8=f_8 \cos\vartheta_8, \quad F_{\eta'}^8=f_8 \sin\vartheta_8,  \\
\label{F0}
&&F_\eta^0=-f_0\sin\vartheta_0, \quad F_{\eta'}^0=f_0\cos\vartheta_0,
\end{eqnarray}
then the relation (\ref{GOf}) is identically satisfied, and the formula (\ref{R2}) takes the form
\begin{equation}
f_0 f_8 \sin (\vartheta_8-\vartheta_0) =\frac{2\sqrt 2}{3}\left(f_\pi^2-f_K^2\right).
\end{equation}
Taking here into account that 
\begin{equation}
f_0 f_8 =F^2\left[ 1+(\hat m+m_s)\frac{a-\delta_M}{2M_0}\right]=f_K^2,
\end{equation}
we arrive to the modified Leutwyler formula \cite{Leutwyler:98}
\begin{equation}
\sin(\vartheta_0-\vartheta_8)=\frac{2\sqrt 2(f_K^2-f_\pi^2)}{3f_K^2}
\end{equation}
from which it is possible to determine the value of the difference $\vartheta_0-\vartheta_8=25^\circ$.

The problem with using the formulas (\ref{F8}) and (\ref{F0}) is that when they are substituted into the linear relations (\ref{LR}) we are forced to conclude that $\vartheta_0=\vartheta_8=\theta$. 

The reason is clear. Both linear (\ref{LRR}) and quadratic (\ref{QR}) relations, when used, imply the rejection of higher-order terms. This means that the formulas (\ref{F8})-(\ref{F0}) also need to be limited to terms of the required precision. In the main text of the paper, we showed how this can be realized.



\begin{thebibliography}{99}
\bibitem{Weinberg:75} S. Weinberg, \emph{The $U(1)$ problem}, Phys. Rev. D {\bf 11} (1975) 3583-3593.
\bibitem{Gross:79} D.\,J. Gross, S.\,B. Treiman, and F. Wilczek, \emph{Light-quark masses and isospin violation}, Phys. Rev. D {\bf 19} (1979) 2188-2196.
\bibitem{Osipov:23} A.\, A. Osipov, \emph{$1/N_c$ Nambu -- Jona-Lasinio model: Electrically charged and strange
    pseudoscalars.} arXiv:hep-ph/2302.14118 (2023).
\bibitem{Nambu:61a} Y. Nambu, G. Jona-Lasinio, \emph{Dynamical model of elementary particles based on an analogy with superconductivity. I}, Phys. Rev. {\bf 122} (1961) 345-358.
\bibitem{Nambu:61b} Y. Nambu, G. Jona-Lasinio, \emph{Dynamical model of elementary particles based on an analogy with superconductivity. II}, Phys. Rev. {\bf 124} (1961) 246-254.
\bibitem{Leutwyler:96a} H. Leutwyler, \emph{Bounds on the light quark masses}, Phys. Lett. B {\bf 374} (1996) 163-168.
\bibitem{Leutwyler:96b} H. Leutwyler, \emph{Implications of $\eta-\eta'$ mixing for the decay $\eta\to 3\pi$}, Phys. Lett. B {\bf 374} (1996) 181-185.
\bibitem{Osipov:21a} A.\,A. Osipov, \emph{Fock-Schwinger method in the case of different masses}, JETP Letters {\bf 113} No.6 (2021) 413-417. 
\bibitem{Osipov:21b} A.\,A. Osipov, \emph{Proper-time method for unequal masses}, Phys. Lett. B {\bf 817} (2021) 136300.
\bibitem{Osipov:21c} A.\,A. Osipov, \emph{Proper-time evaluation of the effective action: Unequal masses in the loop}, Phys. Rev. D {\bf 104} No.10 (2021) 105019.
\bibitem{Gan:22} L. Gan,  B. Kubis, E. Passemar and S. Tulin, \emph{Precision tests of fundamental physics with  $\eta$ and $\eta'$ mesons}, Phys. Rep. {\bf 945} (2022) 1-105.
\bibitem{Osipov:22b} A.\,A. Osipov, \emph{$\pi^0$-$\eta$-$\eta'$ mixing in the theory with four-quark interactions}, 
JETP Letters {\bf 115} (2022) 371-376. 
\bibitem{Taron:97} R. Herrera-Sikl\' ody, J.\,I. Latorre, P. Pascual, J. Taron, \emph{Chiral effective lagrangian in the large-$N_c$ limit: the nonet case}, Nucl. Phys. B {\bf 497} (1997) 345-386.
\bibitem{Kaiser:00} R. Kaiser and H. Leutwyler, \emph{Large $N_c$ in chiral perturbation theory}, Eur. Phys. J. C {\bf 17} (2000) 623-649. 
\bibitem{Goity:02} J. L. Goity, A. M. Bernstein and B. R. Holstein, \emph{Decay $\pi^0\to\gamma\gamma$ to next to leading order in chiral perturbation theory}, Phys. Rev. D {\bf 66} (2002) 076014.
\bibitem{Bickert:20} P. Bickert and S. Scherer, \emph{Two-photon decays and transition form factors of $\pi^0$, $\eta$, and $\eta'$ in large-$N_c$ chiral perturbation theory}, Phys. Rev. D {\bf 102} (2020) 074019.
\bibitem{Volkov:84} M.\,K. Volkov, \emph{Meson Lagrangians in a superconductor quark model}, Ann. of Phys. {\bf 157}  (1984) 282-303.
\bibitem{Wadia:85} A. Dhar, R. Shankar, S.\,R. Wadia, \emph{Nambu--Jona-Lasinio--type effective Lagrangian: Anomalies and nonlinear Lagrangian of low-energy, large-$N$ QCD}, Phys. Rev. D {\bf 31} (1985) 3256-3267.
\bibitem{Volkov:86} M.\,K. Volkov, \emph{Low energy physics of mesons in the superconducting quark model}, PEPAN {\bf 17} (1986) 433-471.
\bibitem{Ebert:86} D. Ebert, H. Reinhardt, \emph{Effective chiral hadron Lagrangian with anomalies and Skyrme terms from quark flavour dynamics}, Nucl. Phys. B {\bf 271} (1986) 188-226.
\bibitem{Osipov:92} V. Bernard, A.\, A. Osipov, U.-G. Mei\ss ner, \emph{Consistent treatment of the bosonized Nambu-Jona-Lasinio model}, Phys. Lett. B {\bf 285} (1992) 119-125.
\bibitem{Bijnens:93} J. Bijnens, C. Bruno and E. de Rafael, \emph{Nambu--Jona-Lasinio-like models and the low-energy effective action of QCD}, Nucl. Phys. B {\bf 390} (1993) 501-541.
\bibitem{Osipov:13} A.\,A. Osipov, B. Hiller, and A.\,H. Blin, \emph{Effective multiquark interactions with explicit breaking of chiral symmetry}, Phys. Rev. D. {\bf 88} (2013) 054032.
\bibitem{Osipov:23b} A.\,A. Osipov, \emph{Gluon anomaly and violation of Zweig's rule}, JETP Letters {\bf 117}, No.  12 (2023) 894-900.
\bibitem{Schechter:93} J. Schechter, A. Subbaraman, H. Weigel, \emph{Effective hadron dynamics: from meson masses to the proton spin puzzle}, Phys. Rev. D {\bf 48} (1993) 339-355.
\bibitem{Moussallam:95} B. Moussallam, \emph{Chiral sum rules for ${\mathcal L}_{(6)}^{WZ}$ parameters and its application to $\pi^0,\eta,\eta'$ decays}, Phys. Rev. D {\bf 51} (1995) 4939-4949.
\bibitem{Feldmann:98} Th. Feldmann, P. Kroll, B. Stech, \emph{Mixing and decay constants of pseudoscalar mesons
}, Phys. Rev. D {\bf 58} (1998) 114006.  
\bibitem{Kroll:05} P. Kroll, \emph{Isospin symmetry breaking through $\pi^0-\eta-\eta'$ mixing}, Mod. Phys. Lett. A {\bf 20} (2005) 2667-2684.
\bibitem{Escribano:05} R. Escribano, J-M. Fr\`ere, \emph{Study of the $\eta$-$\eta'$ system in the two mixing angle scheme}, JHEP 0506:029 (2005). 
\bibitem{Leutwyler:98} H. Leutwyler, \emph{On the $1/N$-expansion in chiral perturbation theory}, Nucl. Phys. B (Proc. Suppl.) {\bf 64} (1998) 223-231.
\bibitem{Klimt:90} S. Klimt, M. Lutz, U, Vogl and W. Weise, \emph{Generalized $SU(3)$ Nambu-- Jona-Lasinio model (I). Mesonic modes}, Nucl. Phys. A {\bf 516} (1990) 429-468.
\bibitem{Hooft:74} G. 't Hooft, \emph{A planar diagram theory for strong interactions}, Nucl. Phys. B {\bf 72} (1974) 461-473.
\bibitem{Veneziano:79} G. Veneziano, \emph{$U(1)$ without instantons}, Nucl. Phys. B {\bf 159} (1979) 213-224.
\bibitem{Witten:79} E. Witten, \emph{Baryons in the $1/N$ expansion}, Nucl. Phys. B {\bf 160} (1979) 57-115.
\bibitem{Witten:79b} E. Witten, \emph{Current algebra theorems for the $U(1)$ Goldstone boson}, Nucl. Phys. B {\bf 156} (1979) 269-283.
\bibitem{Witten:80} E. Witten, \emph{Large N Chiral Dynamics}, Ann. of Phys. {\bf 128} (1980) 363-375.
\bibitem{Veneziano:80} P. Di Vecchia and G. Veneziano, \emph{Chiral dynamics in the large N limit}, Nucl. Phys. B {\bf 171} (1980) 253-272.
\bibitem{Trahern:80} C. Rosenzweig, J. Schechter and G. Trahern, \emph{Is the effective Lagrangian for quantum chromodinamics a $\sigma$ model?}, Phys. Rev. D {\bf 21} (1980) 3388-3392.
\bibitem{Ohta:80} K. Kawarabayashi and N. Ohta, \emph{The $\eta$ problem in the large-N limit: Effective Lagrangian approach}, Nucl. Phys. B {\bf 175} (1980) 477-492.
\bibitem{Ohta:81} K. Kawarabayashi and N. Ohta, \emph{On the partial conservation of the U(1) current},  Prog.  Theor. Phys. {\bf 66} (1981) 1709-1802.
\bibitem{Kordov:21} Z.\,R. Kordov, R. Horsley, W. Kamleh, Z. Koumi, Y. Nakamura, H. Perlt, P.\,E.\,L. Rakow, G. Schierholz, H. St\"uben, R.\,D.\,Young, and J.\,M.\,Zanotti (CSSM/QCDSF/ UKQCD Collaboration) \emph{State mixing and masses of the $\pi^0$, $\eta$ and $\eta'$ mesons from $n_f =1+1+1$ lattice QCD + QED}, Phys. Rev. D {\bf 104} (2021) 114514.
\bibitem{Bramon:97} A. Bramon, R. Escribano, M. Scadron, \emph{Mixing of $\eta-\eta'$ mesons in $J/\psi$ decays into a vector and pseudoscalar meson}, Phys. Lett. B {\bf 403} (1997) 339-343.
\bibitem{Bramon:99} A. Bramon, R. Escribano, M. Scadron, \emph{The $\eta-\eta'$ mixing angle revisited}, Eur. Phys. J. C {\bf 7} (1999) 271-278.
\bibitem{Gasser:85} J. Gasser, H. Leutwyler, \emph{Chiral perturbation theory: Expansions in the mass of the strange quark}, Nucl. Phys. B {\bf 250} (1985) 465-516.
\bibitem{Escribano:16} R. Escribano, S. Gonz\` alez-Sol\'\i s, P. Masjuan, and P. Sanchez-Puertas, \emph{The $\eta'$ transition form factor from space- and time-like experimental data}, Phys. Rev. D {\bf 94} (2016) 054033.
\bibitem{ETM:18} K. Ottnad and C. Urbach, (ETMCollaboration), \emph{ Flavor-singlet meson decay constants from $N_f = 2+1+1$ twisted mass lattice QCD}, Phys. Rev. D {\bf 97}, (2018) 054508.
\bibitem{Teryaev:13} Y. Klopot, A. Oganesian, and O. Teryaev, \emph{Transition form factors and mixing of pseudoscalar mesons from anomaly sum rule}, Phys. Rev. D {\bf 87}, 036013 (2013). 
\bibitem{Klevansky:92} S.\,P. Klevansky, \emph{The Nambu -- Jona-Lasinio model of quantum chromodynamics}, Rev. Mod. Phys. {\bf 64} (1992) 649-708.
\bibitem{Cvetic:97} G. Cvetic, \emph{Regularization at the next-to-leading order in the top-mode standard model without gauge bosons}, Annals of Physics {\bf 255} (1997) 165-203.
\bibitem{PDG} R.\,L. Workman et al. (Particle Data Group), \emph{Review of particle physics}, Prog. Theor. Exp. Phys. {\bf 2022}, 083C01 (2022).
\bibitem{Bickert:17} P. Bickert, P. Masjuan, and S. Scherer, \emph{$\eta$-$\eta'$ mixing in large-$N_c$ chiral perturbation theory}, Phys. Rev. D {\bf 95}, (2017) 054023.
\bibitem{Kaplan:86} D.\,B. Kaplan and A.\,V. Manohar, \emph{Current-mass ratios of the light quarks}, Phys. Rev. Lett. {\bf 56} (1986) 2004-2007. 
\end{thebibliography}
\end{document}